\documentclass[aps,pra,twocolumn
,groupedaddress
,superscriptaddress
,amsfonts,
citeautoscript,
a4paper]{revtex4-2}
\usepackage[T1]{fontenc}
\usepackage[latin9]{inputenc}
\usepackage[a4paper]{geometry}
\usepackage{bm}
\usepackage{amsmath}
\usepackage{amssymb}
\usepackage{graphicx}
\usepackage{babel}
\usepackage{lipsum}
\usepackage{subfig}
\usepackage{xcolor}
\usepackage{simpler-wick}
\usepackage{orcidlink}

\begin{document}
\title{Application of Madelung Hydrodynamics to Plasmonics and Nonlinear Optics in Two-Dimensional
		Materials}

\author{Sim\~{a}o S. Cardoso\, \orcidlink{0009-0009-6573-2185}}
\affiliation{Centro de F\'{\i}sica (CF-UM-UP) and Departamento de F\'{\i}sica, Universidade do Minho, P-4710-057 Braga, Portugal}

\author{A. J. Chaves\,\orcidlink{0000-0003-1381-8568}}
\affiliation{Department of Physics, Aeronautics Institute of Technology, 12228-900, So Jos dos Campos, SP, Brazil}
\affiliation{POLIMA---Center for Polariton-driven Light--Matter Interactions, University of Southern Denmark, Campusvej 55, DK-5230 Odense M, Denmark}

\author{N. Asger Mortensen\,\orcidlink{0000-0001-7936-6264}}
\affiliation{POLIMA---Center for Polariton-driven Light--Matter Interactions, University of Southern Denmark, Campusvej 55, DK-5230 Odense M, Denmark}
\affiliation{Danish Institute for Advanced Study, University of Southern Denmark, Campusvej 55, DK-5230 Odense M, Denmark}

\author{N.~M.~R.~Peres\,\orcidlink{0000-0002-7928-8005}}
\email{peres@fisica.uminho.pt}
\affiliation{Centro de F\'{\i}sica (CF-UM-UP) and Departamento de F\'{\i}sica, Universidade do Minho, P-4710-057 Braga, Portugal}
\affiliation{POLIMA---Center for Polariton-driven Light--Matter Interactions, University of Southern Denmark, Campusvej 55, DK-5230 Odense M, Denmark}
\affiliation{International Iberian Nanotechnology Laboratory (INL), Av Mestre Jos\'e Veiga, 4715-330 Braga, Portugal}

\begin{abstract}
This paper explores the application of Madelung hydrodynamic models to study two-dimensional electron gases, with a focus on nonlocal plasmonics and nonlinear optics. We begin by reviewing the derivation of the Madelung equations. Using the Madelung equations in conjunction with Poisson's equation, we calculate the spectrum of magnetoplasmons and the magneto-optical conductivity in the electrostatic regime, incorporating nonlocal corrections due to the Fermi pressure.
In the absence of a magnetic field, we analyze nonlinear and nonlocal second-harmonic generation, demonstrating how plasmon excitation enhances this process. We further discuss the emergence of self-modulation phenomena driven by nonlinearity, leading to the renormalization of the plasmon dispersion. Notably, we show that nonlinearity amplifies nonlocal effects and, leveraging the hydrodynamic formalism, derive a simple analytic expression for the renormalized spectra.
\end{abstract}
\maketitle

\section{Introduction}

The quantum-hydrodynamic model has emerged as a versatile and powerful framework for investigating diverse physical phenomena, enabling the exploration of numerous properties of matter and phenomena in the general area of light-matter interactions.
Its applications range from plasmonics~\cite{Eguiluz1976,Ciraci2013,Yan2015,Ciraci2016,Baghramyan2021} to nonlinear optics~\cite{Corvi1986,Brevet1997,Ginzburg2015,Zayats2018}. The model has also proven valuable in transistor design~\cite{Crabb2021}, the study of two-dimensional materials~\cite{Fritz2009,Chaves2017,Lucas2018,Fritz2022}, and semiconductor optics~\cite{Asger2018}. This wide-ranging applicability underscores the model's importance in modern quantum physics and its potential for future discoveries.

The origin of the quantum-hydrodynamic model can be traced back to the early days of quantum mechanics, with roots in both the statistical interpretation of the wave function and the hydrodynamic formulation of quantum mechanics. In 1926, Born proposed his groundbreaking statistical interpretation of the wave function~\cite{Born1926,Pais1982,Bernstein2005}.
Building on this foundation, Madelung demonstrated in 1927 that the single-particle time-dependent Schr{\"o}dinger equation could be recast in a hydrodynamic form~\cite{Madelung1927}, establishing a fascinating connection between quantum mechanics and the already well-established formalism of fluid dynamics.

The Madelung equations~\cite{Madelung1927}, which bear a striking resemblance to the continuity and Euler equations of fluid dynamics, initially lacked many-body effects such as statistical pressure~\cite{Bonitz2018}. However, the framework has since evolved to incorporate these crucial elements, giving rise to the quantum-hydrodynamic model. This enhanced model includes the Fermi pressure and other many-body contributions, such as exchange and correlation terms~\cite{Ding2017}. Interestingly, the Madelung transformation can be applied in reverse too. Starting from the hydrodynamic model, one can derive an effective Schr{\"o}dinger equation~\cite{Manfredi2001,Alves2020,Alves2022}. This approach opens up new avenues for solving complex quantum systems using well-established
numerical methods.

The dynamics of electronic fluids under the combined influence of electric and magnetic fields often require investigation~\cite{Fetter1985,Cui1991,Halperin1992,Dulikravich1997,Sano2021}.
These fields induce collective excitations in the electronic fluid, necessitating a self-consistent determination of the fields. In such cases, the hydrodynamic model must be solved in conjunction with Maxwell's equations. However, in certain well-defined limits within plasmonics~\cite{Barnes2016}, the non-retarded approximation~\cite{Locarno2023}
can be employed. Under this approximation, Poisson's equation is used in place of Maxwell's equations, simplifying the calculations while maintaining accuracy for specific scenarios.

A notable limitation of the original Madelung equations is their inability to account for many-body effects~\cite{Khan2014}. This shortcoming can be addressed in two ways. First, the original approach can be reformulated using a many-body wave function~\cite{Janossy1968,Khan2014}. Alternatively,
the Madelung equations can be derived from the moments of the Wigner distribution function~\cite{Manfredi2001}. This latter approach naturally incorporates the statistical pressure, the Bohm term, and additional contributions arising from the energy potential function~\cite{Manfredi2001}. Both methods effectively extend the Madelung formalism to account for many-body quantum phenomena.

This paper is organized as follows: in Sec.~\ref{sec:The-Madelung-transformation} we briefly review the derivation of Madelung equations. In Sec.~\ref{sec:Applications} four applications of Madelung hydrodynamics are given. Finally, in Sec.~\ref{sec:Conclusions} we offer our conclusions. A set of Appendices close the work.

\section{The Madelung transformation\protect}
\label{sec:The-Madelung-transformation}

The time-dependent Schr{\"o}dinger equation reads 
\begin{equation}
i\frac{\partial\Psi(\mathbf{r},t)}{dt}=H\Psi(\mathbf{r},t),
\end{equation}
where $H$ is the Hamiltonian of the system and $\Psi(\mathbf{r},t)$ is the wave function, in general, a complex-valued field. The Madelung transformation is given by the following wave function
\begin{equation}
	\Psi(\mathbf{r},t)=\sqrt{n(\mathbf{r},t)}\exp[iS(\mathbf{r},t)],
\end{equation}
where both $n(\mathbf{r},t)$ and $S(\mathbf{r},t)$ are real-valued functions, and substituting into Schr{\"o}dinger's equation, we obtain two new equations (resulting from separating the real and imaginary parts) 
\begin{subequations}
    \begin{equation}
	\frac{\partial n}{\partial t}+\bm{\nabla}\cdot(n\mathbf{v})=0,\label{eq:conti}
\end{equation}
and
\begin{equation}
	\frac{\partial\mathbf{v}}{\partial t}+\frac{1}{2}\nabla\mathbf{v}^{2}=-\frac{1}{m}\bm{\nabla}U+\frac{\hbar^{2}}{2m^{2}}\bm{\nabla}\left(\frac{1}{\sqrt{n}}\nabla^{2}\sqrt{n}\right),\label{eq:Euler}
\end{equation}
\end{subequations}
where  $\mathbf{v}\equiv \hbar \boldsymbol{\nabla}S/m$, $m$ being the mass of the particle, and $U(\mathbf{r},t)$ represents the potential energy landscape through which the particle moves. Clearly, Eq.~(\ref{eq:conti}) has the form of the continuity equation and Eq.~(\ref{eq:Euler}) has the form of the Euler equation describing the motion of a fluid of velocity $\mathbf{v}$. The term $-m^{-1}\bm{\nabla}U$ accounts for the forces applied to the particle. The last term in Eq.~(\ref{eq:Euler}) does not have a classical analog, being of purely quantum mechanical origin; indeed it goes to zero if $\hbar\rightarrow0$. 
This term is commonly termed the quantum potential or Bohm potential, and it gives rise to spatial dispersion, i.e., physical observables acquire a $\mathbf{k}$-dependence in Fourier space.
When $U(\mathbf{r},t)$ represents a self-consistent potential, $V_\mathrm{sc}(\mathbf{r},t)$, of
electrostatic nature, the solution of the previous two equations must be supplemented by Poisson's equation 
\begin{equation}
	\nabla^{2}V_\mathrm{sc}(\mathbf{r},t)=-\frac{q}{\epsilon_{0}}n(\mathbf{r},t),\label{eq:Poisson}
\end{equation}
where $q$ is the charge of the particles (for electrons we have $q=-e<0$, where $e$ is the elementary charge) and $\epsilon_{0}$ is the vacuum dielectric constant. The use of Poisson's equation is valid in the non-retarded limit. If the inclusion of retardation is necessary then Poisson's equation should be replaced by Maxwell's equations.

The Euler equation derived above includes no many-body effects. This is because it was derived from the single-particle Schr{\"o}dinger's equation. The inclusion of many-body effects can be done phenomenologically, adding by hand the required many-body terms or more formally, by starting with either a many-body wavefunction, a many-body Hamiltonian, or the Wigner distribution function~\cite{Khan2014} (see Appendix \ref{sec:Wigner-function-and}).
 In the many-body approach, and considering a system of $N$ particles, the wave function of the system is written in the form 
\begin{align}
	\Psi(\mathbf{r}_{1}\ldots\mathbf{r}_{N},t) & =\sqrt{n(\mathbf{r}_{1}\ldots\mathbf{r}_{N},t)}\exp[S(\mathbf{r}_{1}\ldots\mathbf{r}_{N},t)]\,.
\end{align}

We can further simplify the wave function assuming that it is possible to write it in factorizable form (Hartree approximation; this factorization is not true in general) as $	\Psi(\mathbf{r}_{1}\ldots\mathbf{r}_{N},t) =\Pi_{i=1}^{N}\sqrt{n_i(\mathbf{r}_{i},t)}\exp[S_i(\mathbf{r}_{i},t)]$.
The Madelung equations also follows from this simplifided form of the many-body wave function.

Either way, at the simplest level, the right-hand-side of the Euler equation above is supplemented by the Fermi pressure term
\begin{equation}
	{\cal \bm{F}}_{F}=-\frac{\bm{\nabla}p_{F}[n(\mathbf{r},t)]}{mn},
\end{equation}
where $p_{F}$ is the Fermi (or statistical) pressure, which depends on the dimensionality of the system and the particles' dispersion relation. In both two-dimensional (2D) and three-dimensional (3D) free-electron systems, the Fermi pressure is taken that of a degenerate electron gas and reads 
\begin{subequations}
\begin{align}
	p_{F} & =\frac{1}{5}\frac{\hbar^{2}n}{m}(3\pi^{2}n)^{2/3},&({\rm 3D\;electron\;gas})\\
	p_{F} & =\pi\frac{\hbar^{2}n^{2}}{2m},&({\rm  2D\;electron\;gas})\\
	p_{F} & =\frac{\hbar v_{F}}{3\pi}(\pi n)^{3/2},&({\rm graphene})
\end{align}\end{subequations}
where $m$ is the electron mass and $v_{F}$ is the Fermi velocity of electrons in graphene. 
In addition to the Bohm potential, the pressure term also contributes to the wavevector dependence of the dielectric function. 
The distinction between the two arises from the power of the wavevector dependence: $k^{2}$ for the Fermi pressure and $k^{4}$ for the Bohm potential.

\section{Applications to plasmonics and nonlinear optics}
\label{sec:Applications}

In this section, we provide four applications of the hydrodynamic model including the effect of quantum corrections. As we will see, this framework is a rather convenient and transparent way to account for nonlocal terms in the response functions.

\subsection{Magnetoplasmons spectrum with nonlocal corrections}

The hydrodynamic formulation of quantum mechanics, briefly outlined above, has proven quite effective in addressing problems such as second-harmonic generation by 2D and 3D electron gases as well as the calculation of the linear-response plasmon dispersion both with and without magnetic field -- magnetoplasmons. In the presence of both an electric and a magnetic field, the Euler equation reads 
\begin{flalign}
\frac{\partial{\cal \bm{V}}}{\partial t}+({\cal \bm{V}}\cdot\bm{\nabla}){\cal \bm{V}}&=\frac{\hbar^{2}}{2m^{2}}\bm{\nabla}\left(\frac{1}{\sqrt{n}}\nabla^{2}\sqrt{n}\right)-\frac{q}{m}\bm{\nabla}V-\nonumber\\&-\frac{q}{m}\frac{\partial\mathbf{A}}{\partial t}+\frac{q}{m}{\cal \bm{V}}\times(\bm{\nabla}\times\mathbf{A}),\label{eq:Euler-equation-magnetic-field}
\end{flalign}
where the electric field reads $\mathbf{E}=-\partial\mathbf{A}/\partial t-\bm{\nabla}V$,
the magnetic field is given by $\mathbf{B}=\bm{\nabla}\times\mathbf{A}$,
and the velocity field is ${\cal \bm{V}}=\mathbf{v}-q\mathbf{A}/m$.
The quantities $V(\mathbf{r},t)$ and $\mathbf{A}(\mathbf{r},t)$ stand for the scalar and vector potentials. Again, the previous equation excludes the Fermi pressure, which will be added by hand. Also, we assume a weak magnetic field allows the use of the same expressions for the Fermi pressure given above. When an external electromagnetic field is added to the system, the magnetic field part of it needs to be included for the sake of the consistent calculation of nonlinear optical properties.

For determining the magnetoplasmon spectrum of an electron gas, we need to solve Eq.~(\ref{eq:Euler-equation-magnetic-field}). As an example, we here consider a 2D electron gas. The force due to the statistical pressure reads 
\begin{equation}
{\cal \bm{F}}_{F}=-\frac{1}{nm}\bm{\nabla}p_{F}=-\frac{\pi\hbar}{2nm^{2}}\bm{\nabla}n^{2}=-\frac{\pi\hbar}{m^{2}}\bm{\nabla}n.
\end{equation}
Treating the problem in the quasi-static approximation see Appendix \ref{App:static}, we have to solve the Euler equation
\begin{equation}
\frac{\partial{\cal \bm{V}}}{\partial t}+({\cal \bm{V}}\cdot\bm{\nabla}){\cal \bm{V}}=-\frac{\pi\hbar}{m^{2}}\bm{\nabla}n-\frac{q}{m}\bm{\nabla}V_\mathrm{sc}+\frac{q}{m}{\cal \bm{V}}\times\mathbf{B},
\end{equation}
together with the continuity equation~(\ref{eq:conti}) and Poisson's equation~(\ref{eq:Poisson}). 
To that end, we assume a time and spatial dependence of the fields in the form $e^{i(\mathbf{r}\cdot\mathbf{k-}\omega t)}$, where both $\mathbf{r}$ and $\mathbf{k}$ are 2D vectors lying in the plane of the 2D electron gas. We then expand them up to linear order in the electric field $\mathbf{E}_\mathrm{sc}=-\bm{\nabla}V_\mathrm{sc}(\mathbf{r},z)$. 
Note that the electromagnetic field exists throughout all space, not just at the surface of the 2D electron gas. The expansion of the fields read
\begin{subequations}
\begin{eqnarray}
n(\mathbf{r},t)=n_{0}+e^{i(\mathbf{r}\cdot\mathbf{k-}\omega t)}n_{1}(\mathbf{k},\omega),\\
\bm{{\cal V}}(\mathbf{r},t)=\bm{{\cal V}}_{0}+e^{i(\mathbf{r}\cdot\mathbf{k-}\omega t)}\bm{{\cal V}}_{1}(\mathbf{k},\omega),
\end{eqnarray}
\end{subequations}
where $n_0$ is the equilibrium two-dimensional electronic density and we consider the absence of drift currents, i.e., $\bm{{\cal V}}_{0}=0$. Thus, to linear order, we have for the continuity and Euler equations
\begin{subequations}
\begin{equation}
\omega n_{1}=n_{0}\mathbf{k}\cdot\bm{{\cal V}}_{1},\label{eq:conti_k_space}
\end{equation}
and
\begin{equation}
\omega\bm{{\cal V}}_{1}=\mathbf{k}\frac{\pi\hbar}{m^{2}}n_{1}+\mathbf{k}\frac{q}{m}V_\mathrm{sc}+i\frac{q}{m}{\cal \bm{V}}_{1}\times\mathbf{B},\label{eq:Euler_k_space}
\end{equation}
while the Poisson's equation reads 
\begin{equation}
\left(\frac{d^{2}}{dz^{2}}-k^{2}\right)V_\mathrm{sc}(\mathbf{k},\omega,z)=-\frac{q}{\epsilon_{0}}\delta(z)n_{1}(\mathbf{k},\omega).\label{eq:Poisson_1D}
\end{equation}
\end{subequations}
The solution of this set of equations at $z=0$ is given by
\begin{equation}
V_\mathrm{sc}(\mathbf{k},\omega,0)=\frac{q}{2k\epsilon_{0}}n_{1}(\mathbf{k},\omega),
\end{equation}
and $n_{1}(\mathbf{k},\omega)$ is given by Eq.~(\ref{eq:conti_k_space}). Replacing $V_\mathrm{sc}(\mathbf{k},\omega,0)$ in Eq.~(\ref{eq:Euler_k_space}), we obtain 
\begin{equation}
\left(\begin{array}{cc}
k_{x}^{2} \frac{\hbar^{2}\pi}{m^{2}}-\frac{\omega^{2}}{n_0}& \frac{k_{x}k_{y}\hbar^{2}\pi}{m^{2}}+i\frac{q\omega B}{n_0 m}\\
\frac{k_{x}k_{y}\hbar^{2}\pi}{m^{2}}-i\frac{q\omega B}{n_0m} & k_{y}^{2}\frac{\hbar^{2}\pi}{m^{2}}-\frac{\omega^{2}}{n_0}
\end{array}\right)\left(\begin{array}{c}
{\cal V}_{1x}\\
{\cal V}_{1y}
\end{array}\right)=0.\label{eq:Eigen_magnetoplasmons}
\end{equation}
This equation has nontrivial solutions if the matrix determinant is zero, which gives 
\begin{equation}
\omega^{2}-\beta^2 k^{2}-ak-\omega_{c}^{2}=0,
\end{equation}
where $\beta^{2}=\hbar^{2}\pi n_{0}/m^{2}$, thus obtaining the magnetoplasmons spectrum given by~\cite{Akbari2013}
\begin{equation}
\omega=\Omega_{\mathbf{k}}=\sqrt{\omega_{c}^{2}+ak+\beta^{2}k^{2}},\label{eq:mag_plas_spectrum}
\end{equation}
 $a=q^{2}n_{0}/(2m\epsilon_{0})$, and $\omega_{c}=\vert qB\vert/m$ is the cyclotron frequency. Since the Fermi momentum of a 2D electron gas is given by $\hbar k_{F}=\hbar\sqrt{2\pi n_{0}}$, the Fermi velocity reads $v_{F}^{2}=\hbar^{2}k_{F}^{2}/m^{2}=2\pi\hbar^{2}n_{0}/m^{2}$. This implies that $\beta$ can be written as 
\begin{equation}
\beta^{2}=\frac{v_{F}^{2}}{2},
\end{equation}
where $\beta$ corresponds to the speed of the first sound in a 2D electron gas~\cite{Mazarella2009}. The previous result is valid when neglecting the collisional frequency, that is, at high frequencies compared to the electron-electron collisions rate $\gamma_{ee}=(k_{B}T)^{2}/(\hbar E_{F})$, where $k_{B}$, $T$, and $E_{F}$ are the Boltzmann constant, the temperature, and the Fermi energy, respectively. For the 3D electron gas we have in this regime $\beta^{2}=3v_{F}^{2}/5$. Interpolation formulas exist between the low and high-frequency regimes~\cite{Halevi1995}. 
For selected electron gases we have~\cite{Halevi1995,Mussot2003,Asger2023} 
\begin{subequations}
\begin{align}
\beta^{2} & =\frac{3\omega/4+i\gamma/2}{\omega+i\gamma}v_{F}^{2},&({\rm 2D\;electron\;gas})\\
\beta^{2} & =\frac{3\omega/5+i\gamma/3}{\omega+i\gamma}v_{F}^{2},&({\rm 3D\;electron\;gas})
\end{align}
\end{subequations}
where $\gamma$ is the total relaxation rate. The high-frequency regime ($\gamma=0$) is obtained from the calculation of the Lindhard function~\cite{Lindhard:1954} and expanding it in powers of the wave vector. If the relaxation rate is finite, the Mermin correction for the dielectric function needs to be used~\cite{Mermin:1970}.

\textcolor{blue}{
\subsection{Magneto-optical conductivity with nonlocal corrections}}

When the external potential $V_\mathrm{ex}$ is employed in Eq.~(\ref{eq:Euler-equation-magnetic-field}) in place of the self-consistent potential, the problem reduces to solving the continuity and Euler equations independently.
We then proceed to derive the linear magneto-optical conductivity for a two-dimensional electron gas. In this framework, the linearized Euler equation for each component is expressed as:
\begin{subequations}
\begin{eqnarray}
\omega{\cal V}_{1x}=k_{x}\frac{\hbar^2\pi}{m^2} n_{1}+k_{x}\frac{q}{m}V_\mathrm{ex}+i\frac{q}{m}{\cal V}_{1y}B,\\
\omega{\cal V}_{1y}=k_{y}\frac{\hbar^2\pi}{m^2} n_{1}+k_{y}\frac{q}{m}V_\mathrm{ex}-i\frac{q}{m}{\cal V}_{1x}B.  
\end{eqnarray}\label{eq:euler_comp}
\end{subequations}
Replacing on this result the first-order electron density found in (\ref{eq:conti_k_space}) and noting that the electric field reads $(E_{x},E_{y})=[-ik_{x}V_\mathrm{ex}(\mathbf{k},\omega,0),-ik_{y}V_\mathrm{ex}(\mathbf{k},\omega,0)]$, we find for each vectorial components of the velocity
\begin{subequations}
\begin{align}
\omega{\cal V}_{1x}&=\frac{k_x\beta^2}{\omega}(k_x{\cal V}_{1x}+k_y{\cal V}_{1y})+\frac{iq}{m}(E_x+\omega_c{\cal V}_{1y}),\\
\omega{\cal V}_{1y}&=\frac{k_y\beta^2}{\omega}(k_x{\cal V}_{1x}+k_y{\cal V}_{1y})+\frac{iq}{m}(E_y-\omega_c{\cal V}_{1x}).  
\end{align}\label{eq:euler_comp}
\end{subequations}
which, after solving for the velocity amplitudes, leads to the solution
\begin{align}
    &\left(\begin{array}{c}
{\cal V}_{1x}\\
{\cal V}_{1y}
\end{array}\right) = \frac{iq}{m D(\mathbf{k},\omega)}\times\\
&\times
    \begin{bmatrix}
        \omega - k_y^2\beta^2/\omega & k_xk_y\beta^2/\omega -i\omega_c \\[2ex]
        k_xk_y\beta^2/\omega +i\omega_c  & \omega -k_x^2\beta^2/\omega
    \end{bmatrix}
    \left(\begin{array}{c}
E_x\\
E_y
\end{array}\right),
\end{align}
where $D(\mathbf{k},\omega) = \omega^2-\omega_c^2-k^2\beta^2$. Defining the current as $\mathbf{J}_{1}=qn_{0}\bm{{\cal V}}_{1}$, we obtain
\begin{align}
&\left(\begin{array}{c}
J_{1x}\\
J_{1y}
\end{array}\right)=\frac{iq^{2}n_{0}}{m D(\mathbf{k},\omega)}\times \\
&\times\left[\begin{array}{cc}
\omega - k_y^2\beta^2/\omega & k_xk_y\beta^2/\omega -i\omega_c \\[2ex]
        k_xk_y\beta^2/\omega +i\omega_c  & \omega -k_x^2\beta^2/\omega
\end{array}\right]\left(\begin{array}{c}
E_x\\
E_y
\end{array}\right).\label{eq:Condutivity_tensor}
\end{align}
The conductivity tensor, with nonlocal corrections included, is now read off from Eq.~(\ref{eq:Condutivity_tensor}) as
\begin{equation}
\overset{\text{\tiny$\leftrightarrow$}}{\sigma}=\frac{iq^{2}n_{0}}{m D(\mathbf{k},\omega)}\left[\begin{array}{cc}
\omega - k_y^2\beta^2/\omega & k_xk_y\beta^2/\omega -i\omega_c \\[2ex]
        k_xk_y\beta^2/\omega +i\omega_c  & \omega -k_x^2\beta^2/\omega
\end{array}\right].
\end{equation}
This result aligns with calculations based on Boltzmann's kinetic equation, provided that nonlocal corrections are neglected~\cite{Aires2011} (see also Ref.~\onlinecite{Andreeva2020}).

\subsection{Second-harmonic generation from a two-dimensional electron gas}

\begin{figure}
\includegraphics[scale=0.2]{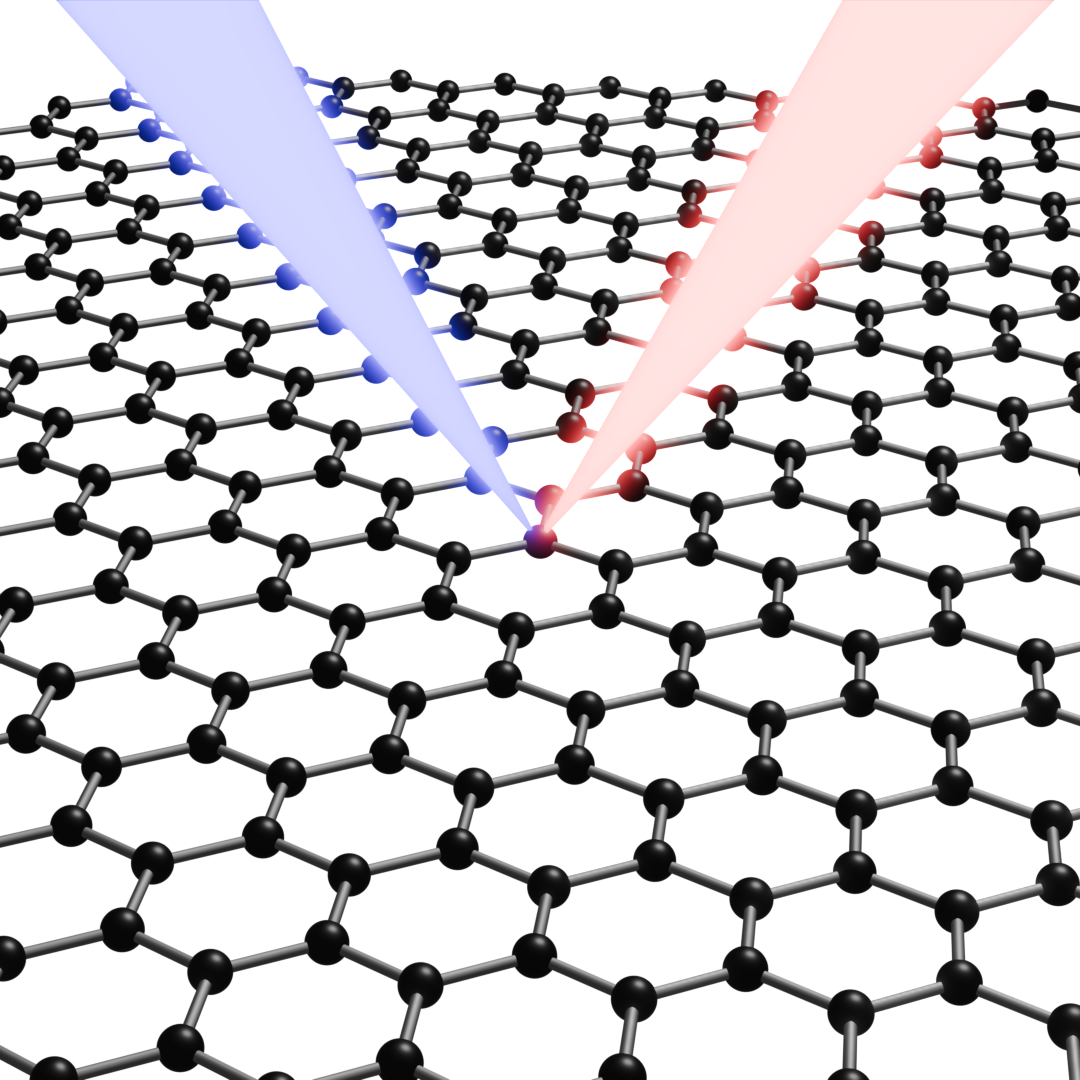}
\caption{Second-harmonic generation. A graphene sheet is illuminated by radiation of frequency $\omega$ (red) and an emerging beam of frequency $2\omega$ is generated (blue). The impinging and outgoing radiation make the same angle as the normal to the graphene sheet.\label{fig:SHG}}
\end{figure}

In the following, we show that the hydrodynamic model facilitates the calculation of nonlinear response functions. As a representative example, we consider a 2D electron gas. To describe second-harmonic generation (SHG), fields must be systematically expanded up to the second order in the external electric field. The system under consideration is illustrated in Fig.~\ref{fig:SHG}.
To account for dissipation, we add a phenomenological damping rate $\gamma$ to the Euler equation~\ref{eq:Euler}
\begin{equation}
	\left(\frac{\partial }{\partial t}+\gamma\right)\mathbf{v} + (\mathbf{v}\cdot\bm{\nabla})\mathbf{v} = \frac{q}{m} \left(\mathbf{E} +\mathbf{v}\times\mathbf{B}\right) ,  
    \label{euler_damped}
\end{equation}
where we considered the zero magnetic field case ${\cal V}=\mathbf{v}$. The expanded fields $n$ and $\mathbf{v}$ up to second order in the external field read
\begin{subequations}    
\begin{align}
	n &=
	n_{0}+n_{1} \nonumber\\
	&=n_{0}+n_{\omega}e^{i(\mathbf{k}\cdot\mathbf{r}-\omega t)}+n_{2\omega}e^{i2(\mathbf{k}\cdot\mathbf{r}-\omega t)}+\mathrm{c.c.},\\
	\mathbf{v}  &=
	\mathbf{v}_{1}+\mathbf{v}_{2}.
\end{align}
\end{subequations}    
where the electric and magnetic fields are defined as
\begin{subequations}    
\begin{align}
\mathbf{E} & =\mathbf{E}_{1}(\mathbf{r})e^{-i\omega t}+\mathbf{E}_{2}(\mathbf{r})e^{-i2\omega t}+\text{ c.c.}
\nonumber\\
&=\frac{\mathbf{E}_{\omega}}{2}e^{i(\mathbf{k}\cdot\mathbf{r}-\omega t)}+\frac{\mathbf{E}_{2\omega}}{2}e^{i2(\mathbf{k}\cdot\mathbf{r}-\omega t)}+\text{ c.c.},\\
\mathbf{B} & =\mathbf{B}_{1}(\mathbf{r})e^{-i\omega t}+\mathbf{B}_{2}(\mathbf{r})e^{-i2\omega t}+\text{ c.c.}
\nonumber\\
&=\frac{\mathbf{B}_{\omega}}{2}e^{i(\mathbf{k}\cdot\mathbf{r}-\omega t)}+\frac{\mathbf{B}_{2\omega}}{2}e^{i2(\mathbf{k}\cdot\mathbf{r}-\omega t)}+\text{ c.c.},
\end{align}
\end{subequations}
where for simplicity the complex conjugated (c.c.) terms have been abbreviated. Substituting these expansions in the continuity equation~(\ref{continuity}) and the Euler equation~(\ref{euler_damped}) and isolating the first and second-order terms in the external electric field, we obtain
\begin{subequations}    
\begin{align}
 & \frac{\partial n_{1}}{\partial t}+n_{0}\bm{\nabla}\cdot\mathbf{v}_{1}+\bm{\nabla}\cdot(n_{1}\mathbf{v}_{1})+n_{0}\bm{\nabla}\cdot\mathbf{v}_{2}=0,\label{continuity}\\
 & \left(\frac{\partial}{\partial t}+\gamma\right)\mathbf{v}_{1}=\frac{q}{m}\mathbf{E},\label{euler_1st}\\
 & \left(\frac{\partial}{\partial t}+\gamma\right)\mathbf{v}_{2}+(\mathbf{v}_{1}\cdot\bm{\nabla})\mathbf{v}_{1}=\frac{q}{m}\mathbf{v}_{1}\times\mathbf{B}.\label{euler_2nd}
\end{align}
\end{subequations}
Solving Eq.~(\ref{euler_1st}), we obtain
\begin{align}
\mathbf{v}_{1} & =i\frac{q}{m}\left(\frac{\mathbf{E}_{1}}{\omega+i\gamma}e^{-i\omega t}+\frac{\mathbf{E}_{2}}{2\omega+i\gamma}e^{-i2\omega t}+\text{ c.c.}\right).\label{v1}
\end{align}
The linear current can be determined using the result~(\ref{v1})
\begin{flalign}
\mathbf{J}_{L}   =qn_{0}\mathbf{v}_{1}  
   =i\frac{e^{2}E_{F}}{\pi\hbar^{2}}\frac{1}{\omega+i\gamma}\frac{\mathbf{E}_{\omega}}{2}e^{-i\omega t}+\nonumber\\+ i\frac{e^{2}E_{F}}{\pi\hbar^{2}}\frac{1}{2\omega+i\gamma}\frac{\mathbf{E}_{2\omega}}{2}e^{-i2\omega t}+\text{ c.c.}
\end{flalign}
Using Ampere's law and Eq.~(\ref{v1}) in Eq.~(\ref{euler_2nd}), we have
\begin{multline}
   \left(\frac{\partial}{\partial t}+\gamma\right)\mathbf{v}_{2}+(\mathbf{v}_{1}\cdot\bm{\nabla})\mathbf{v}_{1}=\frac{q}{m}\mathbf{v}_{1}\\\times\left[-\bm{\nabla}\times\int dt\left(\mathbf{E}_{1}e^{-i\omega t}+\mathbf{E}_{2}e^{-i2\omega t}+\text{ c.c.}\right)\right].
\end{multline}
Direct integration of the previous equation, using the vectorial identity
\begin{equation}
\mathbf{E}_{1}\times(\bm{\nabla}\times\mathbf{E}_{1})=\frac{1}{2}\bm{\nabla}(\mathbf{E}_{1}\cdot\mathbf{E}_{1})-(\mathbf{E}_{1}\cdot\bm{\nabla})\mathbf{E}_{1},
\end{equation}
leads to
\begin{subequations}
\begin{equation}
 \mathbf{v}_{2}=
 \frac{iq^{2}}{m^{2}}\frac{1}{\omega+i\gamma} \left( T_1e^{-i\omega t}+T_2 e^{-2i\omega t} \right)+\mathrm{c.c.}\label{eq:v2}
 \end{equation}
where we defined the auxiliary functions
\begin{equation}
    T_2\equiv\frac{1}{(2\omega+i\gamma)\omega}\left[\frac{-i\gamma}{ \omega+i\gamma}(\mathbf{E}_{1}\cdot\bm{\nabla})\mathbf{E}_{1}+\frac{1}{2}\bm{\nabla}(\mathbf{E}_{1}\cdot\mathbf{E}_{1})\right],
\end{equation}
and
\begin{flalign}
    T_1=-\frac{1}{\omega-i\gamma}\left[\frac{1}{2\omega+i\gamma}\left((\mathbf{E}_{1}^{*}\cdot\bm{\nabla})\mathbf{E}_{2}-  (\mathbf{E}_{2}\cdot\bm{\nabla})\mathbf{E}_{1}^{*}\right) \right. \nonumber\\ \left.-\frac{1}{2\omega} \mathbf{E}_{1}^{*}\times(\bm{\nabla}\times\mathbf{E}_{2}) \right]  -\frac{1}{\omega}\frac{1}{2\omega+i\gamma}\mathbf{E}_{2}\times(\bm{\nabla}\times\mathbf{E}_{1}^{*}).
\end{flalign} 
\end{subequations}

Calculating now the electron density amplitudes from Eq.~(\ref{continuity}), we obtain for the first harmonic
\begin{multline}
n_1=\frac{qn_{0}}{m}\frac{1}{\omega}\frac{1}{\omega+i\gamma}\bm{\nabla}\cdot\mathbf{E}_{1}+\frac{q}{m}n_{2\omega}\frac{1}{\omega}\frac{1}{\omega-i\gamma}\bm{\nabla}\cdot\mathbf{E}_{1}^{*}\\+ \frac{q}{m}n_{\omega}^{*}\frac{1}{\omega}\frac{1}{2\omega+i\gamma}\bm{\nabla}\cdot\mathbf{E}_{2}+\text{ c.c.}, 
\end{multline}
and for the second harmonic ($\mp i2\omega t$)
\begin{align}
n_{2}=\frac{q}{m}\frac{1}{2\omega}\left(\frac{n_{0}}{2\omega+i\gamma}\bm{\nabla}\cdot\mathbf{E}_{2}+  \frac{n_{\omega}}{\omega+i\gamma}\bm{\nabla}\cdot\mathbf{E}_{1}\right)+\text{ c.c.}\label{eq:n2_JNL}
\end{align}
The nonlinear current follows from
\begin{equation}
\mathbf{J}_\mathrm{NL}=qn_{0}\mathbf{v}_{2}+qn_{1}\mathbf{v}_{1}, \label{eq:JNL}
\end{equation}
with $n_1$ given by Eq.~(\ref{eq:n1}), $\mathbf{v}_1$ and $\mathbf{v}_2$ given by Eqs.~(\ref{v1}) and (\ref{eq:v2}), respectively. An explicit expression for $\mathbf{J}_\mathrm{NL} $ is given in Appendix \ref{app:JNL}. 
Making $\gamma=0$ in and using the following vector calculus identity
\begin{align}
&\mathbf{E}_{1}^{*}\times(\bm{\nabla}\times\mathbf{E}_{2})+\mathbf{E}_{2}\times(\bm{\nabla}\times\mathbf{E}_{1}^{*})
\nonumber\\
&=\bm{\nabla}(\mathbf{E}_{1}^{*}\cdot\mathbf{E}_{2})-(\mathbf{E}_{1}^{*}\cdot\bm{\nabla})\mathbf{E}_{2}-(\mathbf{E}_{2}\cdot\bm{\nabla})\mathbf{E}_{1}^{*},
\end{align}
and assuming that the incoming electric field is associated with a wave vector aligned along the $x$-direction, $\mathbf{k}=(k_{x},0,0)$, the nonlinear current becomes
\begin{multline}
\mathbf{J}_{\mathrm{NL},x}  =\frac{3}{2}k_{x}\frac{e^{3}v_{F}^{2}}{4\pi\hbar^{2}\omega^{3}}E_{\omega,x}^{2}e^{-i2\omega t}
\\
-\frac{3}{2}k_{x}\frac{e^{3}v_{F}^{2}}{4\pi\hbar^{2}\omega^{3}}E_{\omega,x}^{*}E_{2\omega,x}e^{-i\omega t}+\text{ c.c.},
\end{multline}
This result agrees with the literature~\cite{Manzoni2015,Cox2018} and where we have used $q=-e<0$, with $e$ the elementary charge.
We stress that in this calculation, the quantity $\mathbf{E}_{\omega}(\mathbf{r})$  is \emph{not} the self-consistent electric field due to plasmonic effects, which is later computed in Sec. \ref{Plasmon-assisted}. For finite $\gamma$ we obtain for the current oscillating at frequency $2\omega$ the result
\begin{align}
\mathbf{J}_{\mathrm{NL},x} & =k_{x}\frac{e^{3}v_{F}^{2}}{4\pi\hbar^{2}}\frac{-\gamma+
	3i\omega}{\left(\gamma-i\omega\right)^{2}(\gamma-2i\omega)\omega}E_{\omega,x}^{2}e^{-i2\omega t}+\text{ c.c.},
	\label{eq:J_NL_x}
\end{align}
that, in the limit $\gamma\rightarrow0$, coincides with previous results using the Boltzmann kinetic approach~\cite{Cox2018}. 
The result for $\mathbf{J}_{\mathrm{NL},x}$, with finite $\gamma$, 
decomposed in its different contributions
is given in
Appendix \ref{app:JNL}.

\begin{figure}[h]
\includegraphics[scale=0.25]{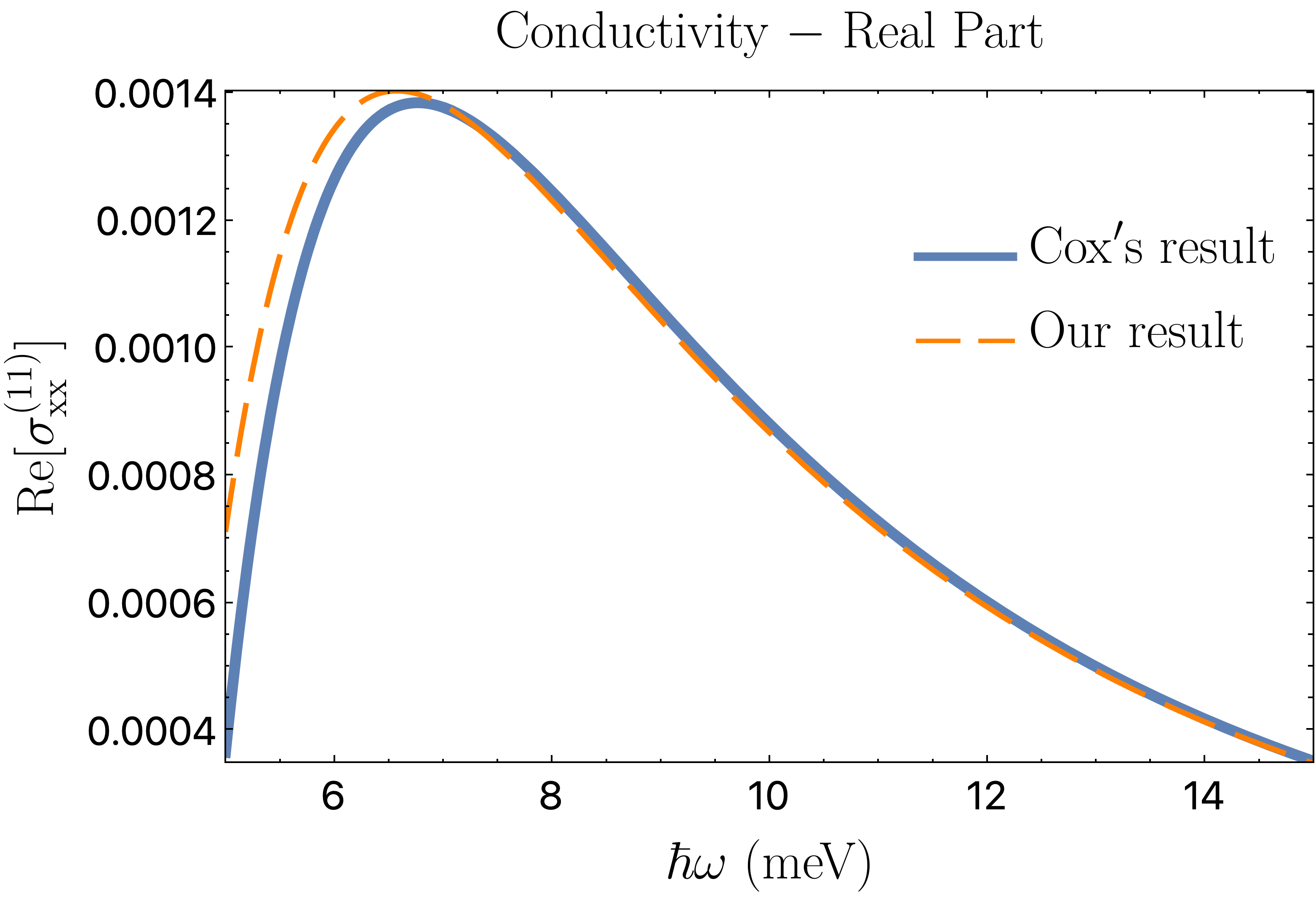}
\includegraphics[scale=0.25]{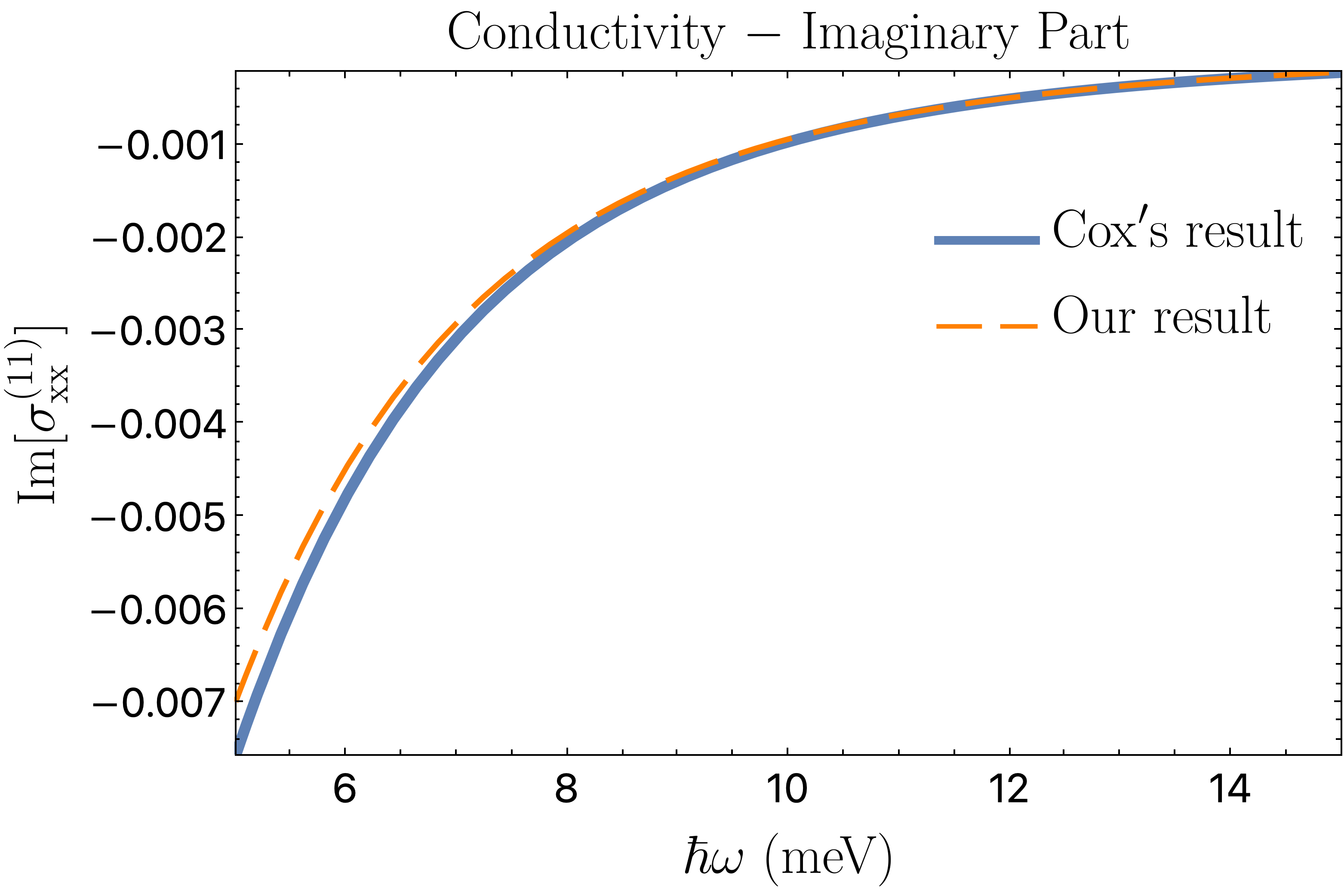}
\caption{Nonlinear optical response for finite $\hbar\gamma=4\;{\rm meV}$ according to Boltzmann kinetic equation~\cite{Cox2018} and hydrodynamic approach (our results). }
\label{fig:Nonlinear-optical-response}
\end{figure}

Our results are compared with the Boltzmann kinetic approach in Fig.~\ref{fig:Nonlinear-optical-response}. 
Although the two analytical expressions
appear different from each other, the numerical results are very similar, as long as we keep the frequency range above the value of the relaxation $\gamma$. 
For $\omega\gg\gamma$, the damping plays no role and the two approaches agree. This is reassuring since the hydrodynamic model is valid precisely in this limit.

\subsubsection{Derivation of the nonlinear current using the polarization}

To assess the correctness of our previous approach, we derive the same results using a more 
common method~\cite{Sipe1980,Scalora2010,Ciraci2012},
where the quantum hydrodynamic equation is expressed as an equation for the polarization $\mathbf{P}$, with the current given by $\mathbf{J}=\dot{\mathbf{P}}$, the dot indicating the partial time derivative. As before, we define the current as $\mathbf{J}=qn\mathbf{v}$. 
The goal is to write the Euler equation for the charged fluid as a function of $\mathbf{J}$. The first step is the integration of the continuity equation written in the form
\begin{equation}
	\dot{n}+\bm{\nabla}\cdot(n\mathbf{v})=\dot{n}+\frac{1}{q}\bm{\nabla}\cdot\mathbf{\dot{P}}=0,
\end{equation}
and upon time integration we find
\begin{equation}
	n=n_{0}-\frac{1}{q}\bm{\nabla}\cdot\mathbf{P}.
\end{equation}
The next step is writing the material derivative of the current, which reads
\begin{equation}
	\frac{d}{dt}\left(qn\mathbf{v}\right)=\frac{\partial}{\partial t}\left(qn\mathbf{v}\right)+(\mathbf{v}\cdot\bm{\nabla})(qn\mathbf{v}).
\end{equation}
This in turn allows the expression of the material derivative of the current as
\begin{equation}
	\frac{d\mathbf{v}}{dt}=\frac{1}{qn}\dot{\mathbf{J}}+\frac{1}{(qn)^{2}}(\mathbf{J}\cdot\bm{\nabla})\mathbf{J}-\frac{\dot{n}}{qn^{2}}\mathbf{J}-\frac{1}{q^{2}n^{3}}(\mathbf{J}\cdot\bm{\nabla}n)\mathbf{J}.
\end{equation}
With this result, the Euler equation~(\ref{euler_damped}) reads
\begin{align}
	\left(\frac{\partial}{\partial t}+\gamma\right)\dot{\mathbf{P}}&=-\frac{1}{qn}(\dot{\mathbf{P}}\cdot\bm{\nabla})\dot{\mathbf{P}}+\frac{\dot{n}}{n}\dot{\mathbf{P}}+\frac{q^{2}n}{m}\mathbf{E}
	\nonumber\\
	&+\frac{q}{m}\dot{\mathbf{P}}\times\mathbf{B}-\frac{q}{m}\bm{\nabla}p_{F}.
\end{align}
What remains is to express $\dot{n}/n$ and $\bm{\nabla}p_{F}$ in terms of the polarization $\mathbf{P}$. The first result
follows from the continuity equation and reads
\begin{equation}
	\frac{\dot{n}}{n}=-\frac{1}{qn_{0}}\bm{\nabla}\cdot\dot{\mathbf{P}}\left(1-\frac{1}{qn_{0}}\bm{\nabla}\cdot\mathbf{P}\right)^{-1}\approx-\frac{1}{qn_{0}}\bm{\nabla}\cdot\dot{\mathbf{P}},
\end{equation}
while the gradient of the pressure is likewise be expressed in terms of the polarization as
\begin{equation}
	-\frac{q}{m}\bm{\nabla}p_{F}=\frac{\pi\hbar^{2}}{m^{2}}\left(n_{0}-\frac{1}{q}\bm{\nabla}\cdot\mathbf{P}\right)\bm{\nabla}\left(\bm{\nabla}\cdot\mathbf{P}\right).
\end{equation}
This latter result, along with the term proportional to $(\mathbf{J}\cdot\bm{\nabla}n)\mathbf{J}$
in the material derivative of the velocity, contributes to nonlocal corrections in orders of $\mathbf{k}$ beyond the linear and are,
therefore, discarded. With these approximations, the equation for the polarization reads
%
\begin{eqnarray}
	\left(\frac{\partial}{\partial t} +\gamma\right)&\dot{\mathbf{P}}\approx-\dfrac{1}{qn_{0}}\left[\left(\dot{\mathbf{P}}\cdot\bm{\nabla}\right)\dot{\mathbf{P}}+    \left(\bm{\nabla}\cdot\dot{\mathbf{P}}\right)\dot{\mathbf{P}}\right]+ \nonumber\\ &+ \dfrac{q^{2}n_{0}}{m}\mathbf{E}-\dfrac{q}{m}\left(\mathbf{E}\left(\bm{\nabla}\cdot\mathbf{P}\right)-\dot{\mathbf{P}}\times\mathbf{B}\right).\label{eq:euler_ciraci}
\end{eqnarray}
%
All the fields entering the previous equation are expanded as
\begin{align}
	\mathbf{A}(\mathbf{r},t) & =\mathbf{A}_{1}(\mathbf{r})e^{-i\omega t}+\mathbf{A}_{2}(\mathbf{r})e^{-i2\omega t}+\text{ c.c.}, \label{expand_fields}
\end{align}
where $\mathbf{A=\mathbf{B}},\mathbf{E},\mathbf{J}$. Replacing this expansions in Eq.~(\ref{eq:euler_ciraci}), we have for the first
harmonic
\begin{equation}
	\mathbf{P}_{1}=-\frac{q^{2}n_{0}}{m}\frac{1}{\omega(\omega+i\gamma)}\mathbf{E}_{1}e^{-i\omega t}+\text{ c.c.},
\end{equation}
and likewise for the second harmonic
\begin{flalign}
	\mathbf{P}_{2}=- \frac{q^{3}n_{0}}{m^{2}D_P(\omega)}     \Bigg[\frac{(\bm{\nabla}\cdot\mathbf{E}_{1})\mathbf{E}_{1}+(\mathbf{E}_{1}\cdot\bm{\nabla})\mathbf{E}_{1}}{\omega+i\gamma}
+\nonumber\\+\frac{\mathbf{E}_{1}(\bm{\nabla}\cdot\mathbf{E}_{1})+\mathbf{E}_{1}\times(\bm{\nabla}\times\mathbf{E}_{1})}{\omega} \Bigg]e^{-i2\omega t}+\text{ c.c.}, \label{eq:P2}
\end{flalign}
with
\begin{equation}
    D_P(\omega)\equiv 2\omega\left(2\omega+i\gamma\right)\left(\omega+i\gamma\right).
\end{equation}

The nonlinear current $\mathbf{J}_2$ can be obtained from Eq.~(\ref{eq:P2}) using that $\mathbf{J}_2=\partial_t \mathbf{P}_2$. 
Assuming a longitudinal electric field oriented along the $x$-direction, we find from the previous expression
\begin{eqnarray}
	J_{2,x}= \frac{e^{3}n_{0}k_x}{m^{2}}\frac{1}{2\omega+i\gamma}\frac{1}{\omega+i\gamma}\Bigg[\frac{2}{\omega+i\gamma}+\frac{1}{\omega}\Bigg]E_{1,x}^{2}e^{-i2\omega t}\nonumber\\ &+\text{ c.c.},
\end{eqnarray}
which, upon algebraic simplification, is identical to Eq.~(\ref{eq:J_NL_x}). Therefore, we have recovered our earlier result.

\subsection{Plasmon-assisted second-harmonic generation in the 2D electron gas}
\label{Plasmon-assisted}

In this section, we show, within the quasi-static approximation, that a self-consistent treatment of the electric potential produces a strong enhancement of the SHG due to resonant plasmon excitation. We take as an example the 2D electron gas. Nonlocal corrections due to the
Fermi pressure is included. We write the continuity and Euler equations
as:
\begin{subequations}\begin{align}
 & 
 \frac{\partial n_\text{ind}}{\partial t}+n_{0}\bm{\nabla}\cdot\mathbf{v}_{1}+\bm{\nabla}\cdot(n_\text{ind}\mathbf{v}_{1})+n_{0}\bm{\nabla} \cdot\mathbf{v}_{2}=0,\label{continuity_SHG}\\
 & \left(\frac{\partial}{\partial t}+\gamma\right)\mathbf{v}_{1}=-\frac{q}{m}\bm{\mathcal{\nabla}}(V_{\text{ext}}+V_\text{ind}),\label{euler_SHG_v1}\\
 & \left(\frac{\partial}{\partial t}+\gamma\right)\mathbf{v}_{2}+(\mathbf{v}_{1}\cdot\bm{\nabla})\mathbf{v}_{1}=0,\label{euler_SHG_v2}
\end{align}\end{subequations}
This set of equations is supplemented by the Poisson equation 
\begin{equation}
\nabla^{2}(V_{\text{ext}}+V_{\text{ind}})=-\frac{q}{\epsilon_{0}}(n_{\text{ind}}-n_{\text{ext}})\delta(z),
\end{equation}
where $n_{\text{ext}}$ is the external charge (located at $z=0$) creating the external potential $V_{\text{ext}}$, while $V_{\text{ind}}$ is the induced potential due to the collective motion of the electron gas, dubbed plasmons. We write the density, external, and induced potentials, respectively, as 
\begin{subequations}
\begin{align}
n_{\text{ind}}(\mathbf{r},t)  =&n_{\omega}e^{i(\mathbf{k}\cdot\mathbf{r}-\omega t)}+n_{2\omega}e^{i2(\mathbf{k}\cdot\mathbf{r}-\omega t)},\\
n_{\text{ext}}(\mathbf{r},t)  =&n_{\omega}^{\text{ext}}e^{i(\mathbf{k}\cdot\mathbf{r}-\omega t)},\\
V_{\text{ext}}(\mathbf{r},z,t) =&\phi_{\omega}^{\text{ext}}(z)e^{i(\mathbf{k}\cdot\mathbf{r}-\omega t)},\\
V_{\text{ind}}(\mathbf{r},z,t)  =&\phi_{\omega}^{\text{ind}}(z)e^{i(\mathbf{k}\cdot\mathbf{r}-\omega t)}\nonumber\\ &+\phi_{2\omega}^{\text{ind}}(z)e^{i2(\mathbf{k}\cdot\mathbf{r}-\omega t)},
\end{align}
\end{subequations}
note that here we are using a different definition than the previous sections, with the density and potentials being complex quantities. 

With these definitions, the Poisson equation splits into two equations, reading
\begin{equation}
\left(\frac{d^{2}}{dz^{2}}-k^{2}\right)\left(\phi_{\omega}^{\text{ext}}(z)+\phi_{\omega}^{\text{ind}}(z)\right)= \frac{q}{\epsilon_{0}}\left(n_{\omega}^{\text{ext}}-n_{\omega}\right)\delta(z),
\end{equation}
and 
\begin{equation}
\left(\frac{d^{2}}{dz^{2}}-4k^{2}\right)\phi_{2\omega}^{\text{ind}}(z)=-\frac{q}{\epsilon_{0}}n_{2\omega}\delta(z).
\end{equation}
The solution of these two equations amounts to the calculation of the Green function and the results are 
\begin{equation}
\phi_{\omega}^{\text{ext}}(z)+\phi_{\omega}^{\text{ind}}(z)=\frac{q}{2k\epsilon_{0}}e^{-k\vert z\vert}(n_{\omega}-n_{\omega}^{\text{ext}}),
\end{equation}
and 
\begin{equation}
\phi_{2\omega}^{\text{ind}}(z)=\frac{q}{4k\epsilon_{0}}e^{-2k\vert z\vert}n_{2\omega}.
\end{equation}
We further define $\phi_{\omega}^{\text{ext}}(0)\equiv\phi_{\omega}^{\text{ext}}$,
$\phi_{\omega}^{\text{ind}}(0)\equiv\phi_{\omega}^{\text{ind}}$, and $\phi_{2\omega}^{\text{ind}}(0)\equiv\phi_{2\omega}^{\text{ind}}$.
The external potential connects to the external density via 
\begin{equation}
\phi_{\omega}^{\text{ext}}=-\frac{q}{2k\epsilon_{0}}n_{\omega}^{\text{ext}},
\end{equation}
and the induced densities are 
\begin{equation}
\phi_{\omega}^{\text{ind}}=\frac{q}{2k\epsilon_{0}}n_{\omega}, \label{ind_1}
\end{equation}
and 
\begin{equation}
\phi_{2\omega}^{\text{ind}}=\frac{q}{4k\epsilon_{0}}n_{2\omega}. \label{ind_2}
\end{equation}
Using these results in Eqs.~(\ref{euler_SHG_v1}) and (\ref{euler_SHG_v2}),
we find
\begin{eqnarray}
\mathbf{v}_{1}=\mathbf{k}\frac{q^{2}}{2\epsilon_{0}mk}\left(\frac{n_{\omega}-n_{\omega}^{\text{ext}}}{\omega+i\gamma}e^{i(\mathbf{k}\cdot\mathbf{r}-\omega t)}+\right. \nonumber \\ \left. +\frac{n_{2\omega}}{2\omega+i\gamma}e^{i2(\mathbf{k}\cdot\mathbf{r}-\omega t)}\right),
\end{eqnarray}
and 
\begin{equation}
\mathbf{v}_{2}=\mathbf{k}\frac{q^{4}}{4\epsilon_{0}^{2}m^{2}}\frac{\left(n_{\omega}-n_{\omega}^{\text{ext}}\right)^{2}}{\left(\omega+i\gamma\right)^{2}(2\omega+i\gamma)}e^{i2(\mathbf{k}\cdot\mathbf{r}-\omega t)}.
\end{equation}

Using these two results in Eq.~(\ref{continuity_SHG}) and grouping
the terms with the same phase, we find the expressions for $n_{\omega}$ and $n_{2\omega}$. With these two results we find from the definition of the linear dielectric function 
\begin{equation}
\phi_{\omega}^{\text{ext}}+\phi_{\omega}^{\text{ind}}=\frac{\phi_{\omega}^{\text{ext}}}{\epsilon(\mathbf{k},\omega)},
\end{equation}
that the dielectric function reads ($\gamma=0$)
\begin{equation}
\epsilon(\mathbf{k},\omega)=1-\omega_{k}^{2}/\omega^{2},
\end{equation}
where $\omega_{k}^{2}=e^{2}n_{0}k/(2\epsilon_{0}m)$ is the plasmon dispersion relation in a 2D electron gas. The second-order electrostatic
amplitude, $\phi_{2\omega}^{\text{ind}}$, reads
\begin{equation}
\phi_{2\omega}^{\text{ind}}=-\frac{3}{8}\frac{e^{3}k^{3}n_{0}}{\epsilon_{0}m^{2}\omega^{4}}\frac{(\phi_{\omega}^{\text{ext}})^{2}}{\epsilon(2\mathbf{k},2\omega)[\epsilon(\mathbf{k},\omega)]^{2}}.\label{eq:phi2w_local}
\end{equation}
This result agrees with a calculation based on the Boltzmann kinetic equation~\cite{Mikhailov2011}. Thus, the second-harmonic field has poles at the solutions of $\epsilon(\mathbf{k},\omega)=0$ and $\epsilon(2\mathbf{k},2\omega)=0$, leading to two peaks in the intensity of the emitted second-harmonic radiation, which is proportional to $(\phi_{\omega}^{\text{ext}})^{2}$.
The vector $\mathbf{k}$ is the in-plane wave vector. Since the external perturbation has the same in-plane wave vector, the SHG is finite only for oblique incidence. Thus, normal incidence originates no SHG, as imposed by symmetry.

So far, our calculation has excluded nonlocal corrections. These can be incorporated by including the contribution of the Fermi pressure in the Euler equation. Apart from this modification, the calculation proceeds exactly as before, yielding
\begin{flalign}
\phi_{2\omega}^{\text{ind}}  =  -\frac{e^{3}k^{3}n_{0}}{2\epsilon_{0}m^{2}}   \frac{1}{(\omega+i\gamma)^{2}} \frac{1}{\hat{\epsilon}(2\mathbf{k},2\omega)}\frac{1}{[\hat{\epsilon}(\mathbf{k},\omega)]^{2}}
\times\nonumber\\  
 \times 
\frac{3\omega^2 - \gamma^2 +\beta^2k^2}{\left(\beta^2k^2 -\omega(\omega+i\gamma)\right)\left(4\beta^2k^2 - 2\omega(2\omega+i\gamma)\right)}
(\phi_{\omega}^{\text{ext}})^{2},\label{eq:SHG_phi}
\end{flalign}
where the denominators can be easily expressed in terms of the dielectric function with damping, as in Eq.~(\ref{eq:phi2w_local}), and which reads 
\begin{equation}
\hat{\epsilon}(\mathbf{k},\omega)=1 - \frac{\omega_k^2}{\omega(\omega+i\gamma) - \beta^2k^2}.
\end{equation}

\begin{figure}
\includegraphics[scale=0.28]{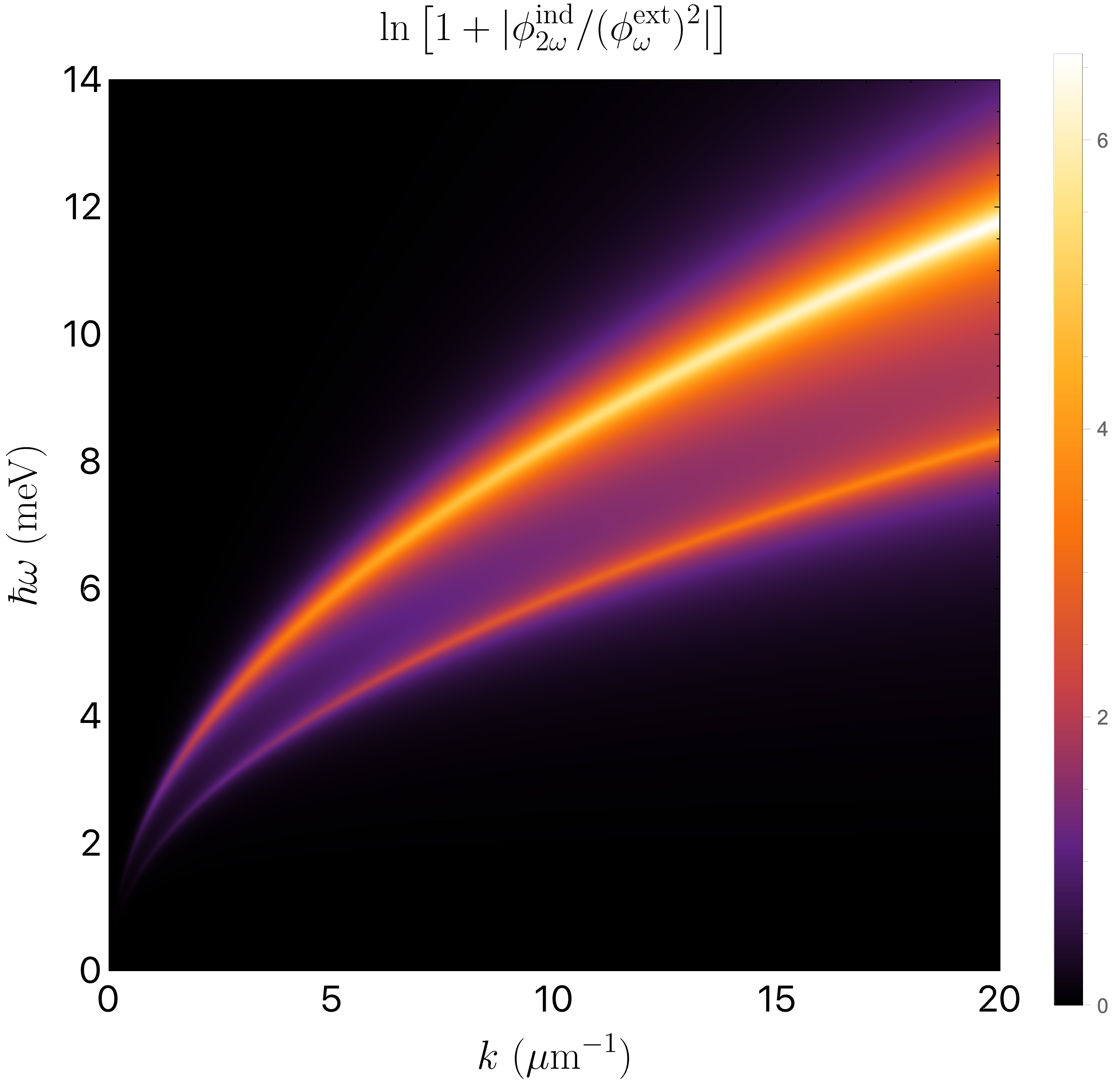}
\caption{Second-harmonic generation assisted by plasmons. The two bright lines correspond to the fundamental and second harmonic. The parameters
are $n_{0}=10^{12}\:{\rm cm^{-2}}$ and $\hbar\gamma=0.25\;{\rm meV}$.
\label{fig:Second-harmonic-generation} }
\end{figure}

The longitudinal excitations of the gas are determined from $\hat{\epsilon}(\mathbf{k},\omega)=0$, implying the spectrum $\omega(k)=\sqrt{\omega_{k}^{2}+\beta^{2}k^{2}}$,
which agrees with Eq.~(\ref{eq:mag_plas_spectrum}) for zero magnetic field. Therefore, the SHG is enhanced by plasmon excitation. The results implied by Eq.~(\ref{eq:SHG_phi}) are depicted in Fig.~\ref{fig:Second-harmonic-generation}.

\subsection{Self-Modulation of SPPs}

A surface-plasmon polariton (SPP) is a coupled oscillation of electromagnetic radiation and charge density at an interface between a metal and a dielectric. Because of their ability to concentrate light in subwavelength volumes, SPPs are of particular interest in nonlinear optics, as they enhance nonlinear optical processes in metallic nanoparticles and on extended metal surfaces. In addition, SPPs are intrinsically nonlinear, capable of altering their properties through self-modulation~\cite{Boyd2014}. This phenomenon arises from the nonlinear interaction between the electric field of the SPP and the medium through which it propagates, resulting in changes to the wave's phase, amplitude, or frequency. A deeper understanding of SPPs and related phenomena paves the way for significant advancements, including the study and design of nonlinear plasmonic systems and metamaterials. In this section, we study the self-modulation effect on the energy of SPPs. For this purpose, we start by expanding the electron velocity and density fields like we have done previously in (\ref{expand_fields}) into the hydrodynamical model, systematically separating harmonics $-i\omega t$ and $-i2\omega t$. Thus, the linear and non-linear hydrodynamical equations become 
\begin{subequations}\begin{align}
    &-i\omega n_1 + n_0\bm{\nabla}\cdot\mathbf{v}_1 + \bm{\nabla}\cdot(n_1^*\mathbf{v}_2) + \bm{\nabla}\cdot(n_2\mathbf{v}_1^*) =0, \label{cont1st} \\
    & - i2\omega n_2 + \bm{\nabla}\cdot(n_1\mathbf{v}_1) + n_0\bm{\nabla}\cdot\mathbf{v}_2 =0, \label{cont2nd} \\
    &-i\omega \mathbf{v}_1 + \left(\mathbf{v}_1^*\cdot\bm{\nabla}\right)\mathbf{v}_2 + \left(\mathbf{v}_2\cdot\bm{\nabla}\right)\mathbf{v}_1^* = \frac{q}{m}\mathbf{E}_1 - b\bm{\nabla}n_1, \label{euler1st}\\
    &-i2\omega \mathbf{v}_2 + \left(\mathbf{v}_1\cdot\bm{\nabla}\right)\mathbf{v}_1 = \frac{q}{m}\mathbf{E}_2 -\frac{\beta^2}{n_0}\bm{\nabla}n_2, \label{euler2nd}
\end{align}\end{subequations}
where $b = \pi\hbar^2/m^2$. 
Referring to the electric potentials given in Eqs.~(\ref{ind_1})--(\ref{ind_2}), the electric fields can be determined by taking their gradients, yielding
\begin{equation}
	\frac{\mathbf{E}_{l\omega}}{2} = -i\mathbf{k}\frac{q}{2\varepsilon_0k}n_{l\omega}, \label{efield}
\end{equation}
where $l=1,2$ indicates the order. Further expanding the fields proportional to $\propto e^{i l \mathbf{k}\cdot\mathbf{r}}$ in Eqs.~(\ref{cont1st})--(\ref{euler2nd}) and replacing the electric fields in Eq.~(\ref{efield}), we get:
\begin{subequations}\begin{align}
    &-\omega n_\omega + n_0(\mathbf{k}\cdot\mathbf{v}_\omega)  + n_\omega^* (\mathbf{k}\cdot\mathbf{v}_{2\omega}) + n_{2\omega}(\mathbf{k}\cdot\mathbf{v}_{\omega}^*) =0,  \label{eq1}\\
    &- \omega n_{2\omega} + n_\omega (\mathbf{k}\cdot\mathbf{v}_\omega) + n_0 (\mathbf{k}\cdot\mathbf{v}_{2\omega})=0 , \label{eq2}\\
    &-\omega \mathbf{v}_\omega + 2\left(\mathbf{v}_\omega^*\cdot\mathbf{k}\right)\mathbf{v}_{2\omega} - \left(\mathbf{v}_{2\omega}\cdot\mathbf{k}\right)\mathbf{v}_\omega^*  +\nonumber\\
    &+ \mathbf{k}\left(\frac{q^2}{2m\varepsilon_0k} + \frac{\beta^2}{n_0}\right) n_\omega =0, \label{eq3}\\
    &-2\omega \mathbf{v}_{2\omega} + \left(\mathbf{v}_\omega \cdot\mathbf{k}\right)\mathbf{v}_\omega   +\mathbf{k}\left(\frac{q^2}{2m\varepsilon_0k} +2\frac{\beta^2}{n_0} \right)n_{2\omega} =0. \label{eq4}
\end{align}\end{subequations}
this system can be simplified due to the realness of the fields $n_\omega,n_{2\omega},\mathbf{v}_{\omega},\mathbf{v}_{2\omega}$ in the absence of losses ($\gamma=0$) and using that $\mathbf{v}_{\omega}$ and $\mathbf{v}_{2\omega}$ are parallel to $\mathbf{k}$:
\begin{subequations}\begin{align}
    & \omega n_\omega - k\left(n_0v_\omega +n_\omega v_{2\omega}+n_{2\omega}v_{\omega} \right) =0,  \label{eq1x}\\
    &  \omega n_{2\omega} -k( n_\omega  v_\omega+ n_0 v_{2\omega})=0 , \label{eq2x}\\
    &\omega v_\omega -  kv_\omega v_{2\omega}
 -  g_k n_\omega =0, \label{eq3x}\\
    &2\omega v_{2\omega} - kv_\omega^2   -g_{2k}n_{2\omega} =0. \label{eq4x}
\end{align}\end{subequations}
where we defined $g_k\equiv \frac{q^2}{2m\varepsilon_0} +\beta^2k/n_0$. To solve this system of equations, we start by isolating $n_{2\omega}$ of Eq.~(\ref{eq2x}) and substituting in Eq.~(\ref{eq4x}), obtaining:
\begin{equation}
2\omega v_{2\omega}-kv_\omega^2 -g_{2k}\frac{k}{\omega}\left(n_\omega v_\omega+n_0v_{2\omega} \right)=0,
\end{equation}
this equation can be solved for $v_{2\omega}$, yielding:
\begin{equation}
    v_{2\omega}=A_k   v_\omega^2 + B_kn_\omega v_\omega , \label{eq:v2s}
\end{equation}
where:
\begin{eqnarray}
    A_k\equiv \left(1-\frac{g_{2k} k n_0}{2\omega^2}   \right)^{-1}\frac{k}{2\omega}, \\
    B_k\equiv\left(1-\frac{g_{2k} k n_0}{2\omega^2}   \right)^{-1}
    \frac{g_{2k}k}{2\omega^2},
\end{eqnarray}
we can substitute Eq.~(\ref{eq:v2s}) in Eq.~(\ref{eq2x}) to obtain $n_{2\omega}$ in terms of $v_\omega$ and $n_\omega$:
\begin{equation}
    n_{2\omega}=\frac{k}{\omega}\left( (B_k n_0+1)n_\omega v_\omega +A_kn_0 v_\omega^2 \right), \label{eq:ns}
\end{equation}
now we substitute Eqs.~(\ref{eq:ns}) and (\ref{eq:v2s}) into (\ref{eq1x}) and (\ref{eq3x}):\begin{subequations}
\begin{eqnarray}
    \omega n_\omega -k\left(n_0v_\omega +n_\omega(B_k n_\omega v_\omega+A_kv_\omega^2)+\right.\nonumber \\ \left.+v_\omega \frac{k}{\omega}\left( (B_k n_0+1)n_\omega v_\omega +A_kn_0 v_\omega^2 \right)\right)=0, \label{eq:vw2}
\end{eqnarray}
\begin{eqnarray}
    \omega v_\omega -kv_\omega \left(B_kn_\omega v_\omega+A_kv_\omega^2 \right) -g_k n_\omega=0, \label{eq:n2w}
\end{eqnarray}\end{subequations}
we can solve Eq.~(\ref{eq:n2w}) for $n_\omega$, obtaining:
\begin{equation}
    n_\omega =  \frac{\omega v_\omega -kA_kv_\omega^3}{   g_k +kB_kv_\omega^2}, \label{eq:nw}
\end{equation}

Substituting Eq.~(\ref{eq:nw}) into (\ref{eq:vw2})  we have:
\begin{equation}
\omega^2 = \frac{q^2 n_0}{2m\varepsilon_0}k+\beta^2 k^2+ kv_\omega^2 F(k,\omega),   \label{eq:final} 
\end{equation}
with:
\begin{flalign}
F(k,\omega)\equiv 
 kn_0B_k +\omega A_k    
    +   \frac{B_k\omega   +A_kg_k}{   g_k +kB_kv_\omega^2}     \left(\omega   -kA_kv_\omega^2 \right)    \nonumber\\  +
      \frac{k}{\omega}\left( (B_k n_0+1)
       \left(\omega   -kA_kv_\omega^2 \right)
    +A_kn_0  \left(g_k +kB_kv_\omega^2 \right) \right).  
\end{flalign}

Equation~(\ref{eq:final}) corresponds to the dispersion relation for SPPs with the nonlinear effects included through the velocity field $v_\omega$, when $v_\omega=0$ we recover the dispersion relation of SPPs with the nonlocal correction due to the Fermi pressure without a magnetic field, as we can see in Eq.~(\ref{eq:mag_plas_spectrum}). 

\begin{figure}[h!]
    \centering
    \includegraphics[width=0.5\textwidth]{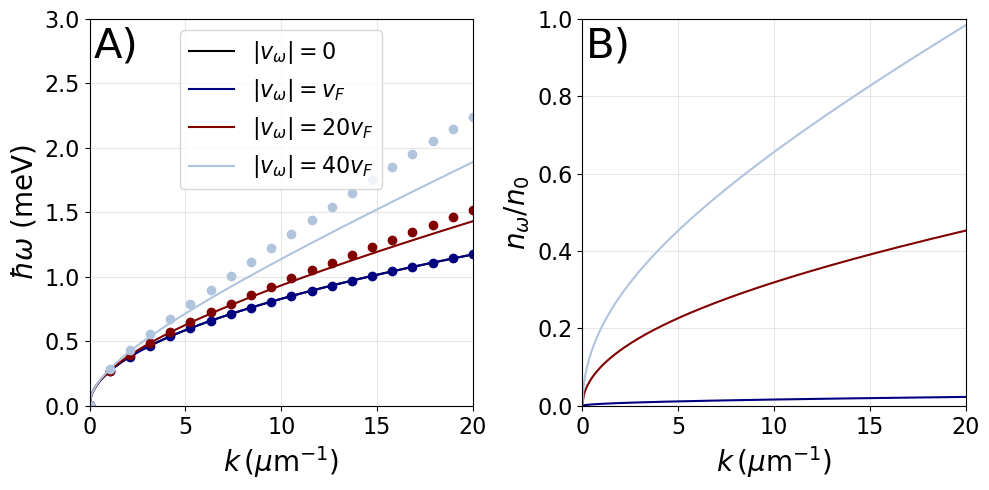}
    \qquad
    \caption{ A) Plasmon-polaritons dispersion relation $\omega(k)$ for different values of $|v_\omega|$ as function of the Fermi velocity $v_F$. B) Charge carrier density fluctuation $n_\omega$ normalized by the charge carrier density $n_0$. Parameters $m=m_0$ and $n_0=10^{14}$\,cm$^{-2}$.}%
    \label{fig:SPPs}
\end{figure}

In Figs.~\ref{fig:SPPs}), we solve iteratively Eq.~(\ref{eq:final}) using as a starting guess in the right-hand side $\omega_0=\sqrt{\frac{q^2 n_0}{2m\varepsilon_0}k+\beta^2 k^2}$ and recursively substituting the left-hand side in the right-hand side for each value of $k$. The dots correspond to a single iteration, while the full lines correspond to the converged value. 
In Fig.~\ref{fig:SPPs}A), we show the dispersion relation, which shows that as the velocity field increases, the frequency increases. For $|v_\omega|=v_F$, the nonlinear effects are negligible. However, there is already a considerable difference for $|v_\omega|=20v_F$. We can see that the dots and the full lines match in this case, which means we can approximate the right-hand side of Eq.~(\ref{eq:final}) to the expression substituting $\omega\rightarrow \omega_0$, but for $|v_\omega|=40v_F$ we see that the straight line and the dots diverges. 

In panel~\ref{fig:SPPs}B), we show the density given by Eq. (\ref{eq:nw}) normalized by the charge carrier density $n_0$,  which shows that the charge density fluctuation $n_\omega$ reaches values close to $n_0$ for high values of $k$ and $|v_\omega|$, meaning that the perturbative approach derived in this section is no more valid. 

In conclusion, we showed how the nonlinear effect self-modulates the plasmon-dispersion. Finally, we note that the nonlinear effects play a similar role to the Fermi pressure, thus being another source of nonlocal contribution for the optical properties.

\section{Conclusions}\label{sec:Conclusions}

In this work, we have applied the hydrodynamic model to various problems as a means of accounting for nonlocal corrections in the computed physical properties of systems. We have demonstrated that quantum hydrodynamic equations can be derived either from the Schr\"{o}dinger equation or from the moments of the Wigner function. Four applications of the quantum-hydrodynamic model were considered, and their results were shown to align with previous literature based on solutions of the Boltzmann kinetic equation.

The hydrodynamic model offers a simple and elegant formalism for addressing the optical properties of condensed matter systems. For instance, our calculation of graphene's nonlinear conductivity can serve as a starting point for discussing second-harmonic generation
in attenuated total reflection experiments. Similarly, the calculation of graphene's magneto-optical conductivity, including nonlocal corrections, provides a foundation for exploring their role in graphene nanostructures under magnetic fields.

The self-consistent field approach to calculating plasmon-assisted second-harmonic generation highlights the simplicity of the method, bypassing the need for a Boltzmann kinetic calculation. 

The proposed results found in this work can be validated experimentally, especially the second harmonic generation. Its measurement is obtained from the ATR configuration (attenuated total reflection) originally proposed by Otto~\cite{otto_1968}, where shining light on a prism, separated from graphene by a dielectric spacer, creates a second-order nonlinear response of the material observed in reflection. On the other hand, measurements on the dispersion relation of plasmons are possible using different techniques, such as ATR or SNOM~\cite{chen_2012} (scanning near-field optical microscope), requiring small distances between the probe, for example, the SNOM tip and the graphene sheet.

\section*{Acknowledgments }

N. M. R. P. acknowledges support from the Portuguese Foundation for
Science and Technology (FCT) in the framework of the Strategic Funding
UIDB/04650/2020, COMPETE 2020, PORTUGAL 2020, FEDER, and through project
PTDC/FIS-MAC/2045/2021. N. M. R. P. also acknowledges the hospitality
of the POLIMA Center and, together with N. A. M., the Independent
Research Fund Denmark (grant no. 2032-00045B) and the Danish National
Research Foundation (Project No. DNRF165).  A.~J.~C. were supported by CNPq (Conselho Nacional de Desenvolvimento Cient\'ifico e Tecnol\'ogico) Grant No. 423423/2021-5, 408144/2022-0, 315408/2021-9 and FAPESP under Grant No.~2022/08086-0 and a CAPES-PrInt scholarship.

\appendix

\section{Connection between the Wigner function and Madelung equations \label{sec:Wigner-function-and}}

In the following, we derive the Madelung equations starting from the Wigner distribution function, which preserves information about both position and momentum while maintaining the interference properties characteristic of quantum mechanics. Assuming a distribution function of the form (in 2D)
\begin{multline}
	f(\mathbf{r},\mathbf{p},t)=\frac{1}{(2\pi)^{2}\hbar^{2}}\sum_{\alpha}P_{\alpha} \\\int d\mathbf{r^{\prime}}\psi_{\alpha}^{\ast} 
	(\mathbf{r}+\mathbf{r}^{\prime}/2,t)\psi_{\alpha}(\mathbf{r}-\mathbf{r}^{\prime}/2,t)e^{i\mathbf{p}\cdot\mathbf{r}^{\prime}/\hbar}, \label{eq:wigner}
\end{multline}
where $\psi_{\alpha}(\mathbf{r},t)$ is a single-particle wave function obeying the time-dependent Schr{\"o}dinger equation. The system is assumed to be in a mixed state of the form 

	\begin{equation}
		\rho(\mathbf{r},\mathbf{r'}, t)=\sum_{\alpha}P_{\alpha}\psi^{\ast}_{\alpha}(\mathbf{r},t)\psi_{\alpha}(\mathbf{r'},t),
	\end{equation}
where the probabilities $P_{\alpha}$ obey the sum rule $\sum_{\alpha}^{N_{s}}P_{\alpha}=1$, where $N_{s}$ is the number of states in the mixture. The time derivative of $f(\mathbf{r},\mathbf{p},t)$ can be obtained using the time-dependent Schr{\"o}dinger equation. Introducing the following two moments of the Wigner distribution  
\begin{subequations}
	\begin{eqnarray}
		n(\mathbf{r},t)=\int d\mathbf{p}\,f(\mathbf{r},\mathbf{p},t), \\
		n(\mathbf{r},t)\mathbf{v}(\mathbf{r},t)=\frac{1}{m}\int d\mathbf{p}\,\mathbf{p}\,f(\mathbf{r},\mathbf{p},t),
	\end{eqnarray} 
\end{subequations}
the continuity and Euler equations follow, with the latter having both the statistical and quantum (Bohm term) pressures included. We derive the continuity equation first. From the time derivative of the Wigner distribution~(\ref{eq:wigner}), using the results presented in Appendix~\ref{App:derivation}, we have:
\begin{equation}
	\frac{\partial}{\partial t} f(\mathbf{r},\mathbf{p},t)=-\frac{1}{m}\bm{\nabla}\cdot[\mathbf{p}f(\mathbf{r},\mathbf{p},t)]+V_{W}(\mathbf{r},\mathbf{p},t),
	\label{eq:A4}
\end{equation}
This result, in the present context, is called the Wigner equation~\cite{Khan2014}.
It closely resembles the Boltzmann kinetic equation, apart from the $V_{W}(\mathbf{r},\mathbf{p},t)$ term.
Upon integration over the momentum $\mathbf{p}$, the term $V_{W}(\mathbf{r},\mathbf{p},t)$ cancels out and the final result reads 
\begin{equation}
	\frac{\partial}{\partial t}n(\mathbf{r},t)+\bm{\nabla}\cdot[n(\mathbf{r},t)\mathbf{v}(\mathbf{r},t)]=0, \label{eq:continuity}
\end{equation}
that is, we have obtained the continuity equation. The calculation of the first moment yields the Euler equation, incorporating both the statistical and quantum pressure terms. To derive the Euler equation, we take the first moment of the Wigner distribution function. Starting with
\begin{multline}
	\mathbf{p}f(\mathbf{r},\mathbf{p},t)=\frac{1}{(2\pi)^{2}\hbar^{2}}\mathbf{p}\sum_{\alpha}P_{\alpha}\\\int d\mathbf{r^{\prime}}\psi_{\alpha}^{\ast}(\mathbf{r}+\mathbf{r}^{\prime}/2,t)
	\psi_{\alpha}(\mathbf{r}-\mathbf{r}^{\prime}/2,t)e^{i\mathbf{p}\cdot\mathbf{r}^{\prime}/\hbar},
\end{multline}
integrating over $\mathbf{p}$ and dividing by $m$, we obtain
\begin{multline}
	\frac{1}{m}\int d\mathbf{p}\mathbf{p}f(\mathbf{r},\mathbf{p},t)=\frac{1}{(2\pi)^{2}\hbar^{2}}\frac{1}{m}\int d\mathbf{p}\mathbf{p}\sum_{\alpha}P_{\alpha}
	\\
	\int d\mathbf{r^{\prime}}
	\psi_{\alpha}^{\ast}(\mathbf{r}+\mathbf{r}^{\prime}/2,t)
	\psi_{\alpha}(\mathbf{r}-\mathbf{r}^{\prime}/2,t)e^{i\mathbf{p}\cdot\mathbf{r}^{\prime}/\hbar}.
\end{multline}
Taking the time derivative of the previous equation and using the Schr\"{o}dinger as we did before, we have 
\begin{equation}
	\frac{\partial[n(\mathbf{r},t)\mathbf{v}(\mathbf{r},t)]}{\partial t}=-\frac{1}{m^{2}}\int d\mathbf{p}\bm{\nabla}\cdot[\mathbf{p}\otimes\mathbf{p}f(\mathbf{r},\mathbf{p},t)],
\end{equation}
which can be written as 
\begin{equation}
	\frac{\partial[n(\mathbf{r},t)\mathbf{v}(\mathbf{r},t)]}{\partial t}+\bm{\nabla}\cdot\bm{\Pi}(\mathbf{r},t)={\cal \bm{P}_{W}}(\mathbf{r},t),\label{eq:Euler_Wigner}
\end{equation}
where the term $\bm{\Pi}$ represents the statistical and quantum (Bohm) pressure, while the term ${\cal P}_{w}$ comes from the contribution of the potential and represents additional force terms. 
We now apply the continuity equation~(\ref{eq:conti}) to the first term of the previous equation. After explicitly performing the time derivative and carrying out some calculations, as shown in Appendix \ref{app:euler}, we obtain
\begin{eqnarray}
	n(\mathbf{r},t)\frac{\partial\mathbf{v}(\mathbf{r},t)}{\partial t}+n(\mathbf{r},t)[\mathbf{v}(\mathbf{r},t)\cdot\bm{\nabla}]\mathbf{v}(\mathbf{r},t)+\nonumber\\+\bm{\nabla}\cdot\left[\bm{\Pi}-n(\mathbf{r},t)\mathbf{v}(\mathbf{r},t)\mathbf{v}(\mathbf{r},t)\right]={\cal \bm{P}_{W}}.
\end{eqnarray}
This corresponds to the Euler equation of the quantum-hydrodynamic model. We can see from the final result that the quantum-pressure terms, $\bm{\Pi}$, appear naturally in the Wigner formulation, which leads to nonlocal corrections. Also, the Wigner distribution accounts for motion in the phase space $(\mathbf{r},\mathbf{p})$, whereas the Madelung approach works only in real space. Thus, in the Wigner approach, quantum phenomena such as interference are naturally incorporated into the formalism.

\section{Quasi-static limit} \label{App:static}

In this section, we show that for SPPs, that are transverse magnetic (TM) modes, in the long wavenumber regime $q\gg \omega/c$, we can associate a scalar potential that obeys the Poisson equation. We will consider that a 2D material located at $z=0$, described by the surface densities and currents $n(\mathbf{r_\parallel,t}), \mathbf{j}(\mathbf{r_\parallel,t})$, is immersed between two semi-infinite dielectrics with dielectric constants $\epsilon_1,\epsilon_2$. The electromagnetic fields satisfies the Maxwell equations:
\begin{subequations}
\begin{eqnarray}
\boldsymbol{\nabla}\cdot \mathbf{E}=n(\mathbf{r}_\parallel,t)\delta(z),\\
\boldsymbol{\nabla}\times \mathbf{E}=-\frac{\partial \mathbf{B}}{\partial t},\\
\boldsymbol{\nabla}\cdot \mathbf{B}=0,\\
\boldsymbol{\nabla}\times \mathbf{B}=\mu_0\mathbf{j}\delta(z)+\frac{1}{c^2}\epsilon(z)\frac{\partial \mathbf{E}}{\partial t},
\end{eqnarray}
\end{subequations}

We start considering SPPs described by the electromagnetic field $(\mathbf{E},\mathbf{B})$piecewise continuous, that are evanescent in the $z$ direction and propagates along the $xy$ directions:
 \begin{equation}
    \mathbf{E}(\mathbf{r},t) = e^{i\mathbf{q}\cdot\mathbf{r}_\parallel-i\omega t}\times\begin{cases}\mathbf{E}_1 e^{-\kappa_1 z}, \mathrm{for}\,\,z>0, \\ \mathbf{E}_2 e^{ +\kappa_2 z}, \mathrm{for}\,\,z<0,  \end{cases}
 \end{equation}
 with $\kappa_i=\sqrt{q^2-\epsilon_1 \omega^2/c^2}$. When $\kappa \gg \epsilon_j \omega^2/c^2$, with $j=1,2$, we can approximate $\kappa \approx q$, therefore:
 \begin{equation}
    \mathbf{E}(\mathbf{r},t) \approx e^{i\mathbf{q}\cdot\mathbf{r}_\parallel-q|z|-i\omega t}\times\begin{cases}\mathbf{E}_1, \mathrm{for}\,\,z>0, \\ \mathbf{E}_2, \mathrm{for}\,\,z<0,  \end{cases}
 \end{equation} 
 that satisfies:
 \begin{equation}
 \boldsymbol{\nabla}\times\mathbf{E}(\mathbf{r},t)=0, \label{eq:curl}
 \end{equation}
 i.e., $\mathbf{E}(\mathbf{r},t)$ admits a scalar potential. To prove (\ref{eq:curl}), we consider, without loss of generality, that $\mathbf{q}=q\mathbf{u}_x$ and therefore $\mathbf{B}=(0,B_y,0)$ which results $\mathbf{E}=(E_x,0,E_z)$. Using that, for $z\neq0$, we have from $(iq,0,-q)\cdot(E_x,0,E_z)=0$, from which follows (\ref{eq:curl}). As the curl of the electric field is approximate zero in the long wavenumber limit, it results that it can be described by a scalar potential that obeys the Poisson equation.

\section{Non-linear current} \label{app:JNL}

Substituting Eqs.~(\ref{eq:n1}), (\ref{v1}) and (\ref{eq:v2}) in Eq. (\ref{eq:JNL}, we obtain:
\begin{widetext}
	\begin{align}
		\mathbf{J}_\mathrm{NL} & =qn_{0}\mathbf{v}_{2}+qn_{1}\mathbf{v}_{1}\nonumber \\
		& =i\frac{q^{3}n_{0}}{m^{2}}\frac{1}{\omega+i\gamma}\Bigg[\frac{1}{2\omega+i\gamma}\left(\frac{1}{\omega+i\gamma}-\frac{1}{\omega}\right)(\mathbf{E}_{1}\cdot\bm{\nabla})\mathbf{E}_{1}+\frac{1}{\omega}\frac{1}{\omega+i\gamma}(\bm{\nabla}\cdot\mathbf{E}_{1})\mathbf{E}_{1}+\frac{1}{2\omega}\frac{1}{2\omega+i\gamma}\bm{\nabla}(\mathbf{E}_{1}\cdot\mathbf{E}_{1})\Bigg]e^{-i2\omega t}\nonumber \\
		& -i\frac{q^{3}n_{0}}{m^{2}}\Bigg[\frac{1}{\omega+i\gamma}\frac{1}{\omega-i\gamma}\frac{1}{2\omega+i\gamma}(\mathbf{E}_{1}^{*}\cdot\bm{\nabla})\mathbf{E}_{2}+\frac{1}{\omega+i\gamma}\frac{1}{\omega-i\gamma}\frac{1}{2\omega+i\gamma}(\mathbf{E}_{2}\cdot\bm{\nabla})\mathbf{E}_{1}^{*}+\frac{1}{\omega+i\gamma}\frac{1}{2\omega}\frac{1}{\omega-i\gamma}\mathbf{E}_{1}^{*}\times(\bm{\nabla}\times\mathbf{E}_{2})\nonumber \\
		& +\frac{1}{\omega+i\gamma}\frac{1}{\omega}\frac{1}{2\omega+i\gamma}\mathbf{E}_{2}\times(\bm{\nabla}\times\mathbf{E}_{1}^{*})-\frac{1}{\omega}\frac{1}{\omega-i\gamma}\frac{1}{2\omega+i\gamma}(\bm{\nabla}\cdot\mathbf{E}_{1}^{*})\mathbf{E}_{2}+\frac{1}{2\omega}\frac{1}{\omega-i\gamma}\frac{1}{2\omega+i\gamma}(\bm{\nabla}\cdot\mathbf{E}_{2})\mathbf{E}_{1}^{*}\Bigg]e^{-i\omega t}+\text{ c.c.}
	\end{align}

\section{Derivation of Eq.  \ref{eq:A4}\label{App:derivation}} 

We start with the  time derivative of Eq. (\ref{eq:wigner}),
that results in
\begin{align}
\frac{\partial}{\partial t}f(\mathbf{r},\mathbf{p},t) & =\frac{1}{(2\pi)^{2}\hbar^{2}}\sum_{\alpha}P_{\alpha}\int d\mathbf{r^{\prime}}\left[\frac{\partial}{\partial t}\psi_{\alpha}^{\ast}(\mathbf{r}+\mathbf{r}^{\prime}/2,t)\right]\psi_{\alpha}(\mathbf{r}-\mathbf{r}^{\prime}/2,t)e^{i\mathbf{p}\cdot\mathbf{r}^{\prime}/\hbar},\nonumber \\
 & +\frac{1}{(2\pi)^{2}\hbar^{2}}\sum_{\alpha}P_{\alpha}\int d\mathbf{r^{\prime}}\psi_{\alpha}^{\ast}(\mathbf{r}+\mathbf{r}^{\prime}/2,t)\frac{\partial}{\partial t}\psi_{\alpha}(\mathbf{r}-\mathbf{r}^{\prime}/2,t)e^{i\mathbf{p}\cdot\mathbf{r}^{\prime}/\hbar},
\end{align}
and using the time-dependent Schr{\"o}dinger equation we have 
\begin{align}
\frac{\partial}{\partial t}f(\mathbf{r},\mathbf{p},t) & =\frac{i}{(2\pi)^{2}\hbar^{2}}\sum_{\alpha}P_{\alpha}\int d\mathbf{r^{\prime}}\left[H\psi_{\alpha}^{\ast}(\mathbf{r}+\mathbf{r}^{\prime}/2,t)\right]\psi_{\alpha}(\mathbf{r}-\mathbf{r}^{\prime}/2,t)e^{i\mathbf{p}\cdot\mathbf{r}^{\prime}/\hbar},\nonumber \\
 & -\frac{i}{(2\pi)^{2}\hbar^{2}}\sum_{\alpha}P_{\alpha}\int d\mathbf{r^{\prime}}\psi_{\alpha}^{\ast}(\mathbf{r}+\mathbf{r}^{\prime}/2,t)H\psi_{\alpha}(\mathbf{r}-\mathbf{r}^{\prime}/2,t)e^{i\mathbf{p}\cdot\mathbf{r}^{\prime}/\hbar}.
\end{align}
The term dependent on the potential energy gives a finite contribution that we will ignore it for the moment and reintroduce it later. We are then left
with the kinetic energy 
\begin{align}
\frac{\partial}{\partial t}f(\mathbf{r},\mathbf{p},t) & =\frac{1}{2im(2\pi)^{2}}\sum_{\alpha}P_{\alpha}\int d\mathbf{r^{\prime}}\left[\nabla^{2}\psi_{\alpha}^{\ast}(\mathbf{r}+\mathbf{r}^{\prime}/2,t)\right]\psi_{\alpha}(\mathbf{r}-\mathbf{r}^{\prime}/2,t)e^{i\mathbf{p}\cdot\mathbf{r}^{\prime}/\hbar}\nonumber \\
 & -\frac{1}{2im(2\pi)^{2}}\sum_{\alpha}P_{\alpha}\int d\mathbf{r^{\prime}}\psi_{\alpha}^{\ast}(\mathbf{r}+\mathbf{r}^{\prime}/2,t)\nabla^{2}\psi_{\alpha}(\mathbf{r}-\mathbf{r}^{\prime}/2,t)e^{i\mathbf{p}\cdot\mathbf{r}^{\prime}/\hbar}.
\end{align}
Next, we change the Laplacian differentiation variable from $\mathbf{r}$
for $\mathbf{r}'$ and use the following identity 
\begin{align}
\bm{\nabla}'\cdot\left[\left(\bm{\nabla}'\psi_{\alpha}^{\ast}(\mathbf{r}+\mathbf{r}^{\prime}/2,t)\right)\psi_{\alpha}(\mathbf{r}-\mathbf{r}^{\prime}/2,t)-\psi_{\alpha}^{\ast}(\mathbf{r}+\mathbf{r}^{\prime}/2,t)\bm{\nabla}'\psi_{\alpha}(\mathbf{r}-\mathbf{r}^{\prime}/2,t)\right] & =\nonumber \\
\left[\nabla'{}^{2}\psi_{\alpha}^{\ast}(\mathbf{r}+\mathbf{r}^{\prime}/2,t)\right]\psi_{\alpha}(\mathbf{r}-\mathbf{r}^{\prime}/2,t)-\psi_{\alpha}^{\ast}(\mathbf{r}+\mathbf{r}^{\prime}/2,t)\nabla'{}^{2}\psi_{\alpha}(\mathbf{r}-\mathbf{r}^{\prime}/2,t) & ,
\end{align}
with this identity, we perform integration by parts leading to 
\begin{alignat}{1}
\frac{\partial}{\partial t}f(\mathbf{r},\mathbf{p},t) & =\frac{4}{2m(2\pi)^{2}\hbar^{2}}\sum_{\alpha}P_{\alpha}\int d\mathbf{r^{\prime}}\mathbf{p}\cdot\left[\left(\bm{\nabla}'\psi_{\alpha}^{\ast}(\mathbf{r}+\mathbf{r}^{\prime}/2,t)\right)\psi_{\alpha}(\mathbf{r}-\mathbf{r}^{\prime}/2,t)
-\right. \nonumber \\  &\left. 
-\psi_{\alpha}^{\ast}(\mathbf{r}+\mathbf{r}^{\prime}/2,t)\bm{\nabla}'\psi_{\alpha}(\mathbf{r}-\mathbf{r}^{\prime}/2,t)\right]e^{i\mathbf{p}\cdot\mathbf{r}^{\prime}/\hbar}\nonumber \\
 & =-\frac{1}{m(2\pi)^{2}\hbar^{2}}\sum_{\alpha}P_{\alpha}\int d\mathbf{r^{\prime}}\mathbf{p}\cdot\bm{\nabla}[\psi_{\alpha}^{\ast}(\mathbf{r}+\mathbf{r}^{\prime}/2,t)\psi_{\alpha}(\mathbf{r}-\mathbf{r}^{\prime}/2,t)]e^{i\mathbf{p}\cdot\mathbf{r}^{\prime}/\hbar}\nonumber \\
 & =-\frac{1}{m}\bm{\nabla}\cdot\mathbf{p}\frac{1}         {(2\pi)^{2}\hbar^{2}}\sum_{\alpha}P_{\alpha}\int d\mathbf{r^{\prime}}\psi_{\alpha}^{\ast}(\mathbf{r}+\mathbf{r}^{\prime}/2,t)\psi_{\alpha}(\mathbf{r}-\mathbf{r}^{\prime}/2,t)e^{i\mathbf{p}\cdot\mathbf{r}^{\prime}/\hbar}\nonumber \\
 & =-\frac{1}{m}\bm{\nabla}\cdot[\mathbf{p}f(\mathbf{r},\mathbf{p},t)]+V_{W}(\mathbf{r},\mathbf{p},t),
\end{alignat}
where in the last line we recovered the contribution from the potential:
\begin{flalign}
V_W(\mathbf{r},\mathbf{p},t)\equiv 
\frac{i}{4\pi^2\hbar^2}\sum_\alpha P_\alpha \int d\mathbf{r}^\prime 
\left(V\left(\mathbf{r}+\mathbf{r}^\prime/2 \right)
-V\left(\mathbf{r}-\mathbf{r}^\prime/2\right)\right)\psi_\alpha^*(\mathbf{r}+\mathbf{r}^\prime/2,t)\psi_\alpha(\mathbf{r}-\mathbf{r}^\prime/2,t)e^{i\mathbf{p}\cdot\mathbf{r}^{\prime}/\hbar}.
\end{flalign}
The derivation of Euler's equation is given in the next Appendix.
\end{widetext}

\section{Derivation of the Euler equation from the Wigner distribution} \label{app:euler}

We start this appendix calculating the time derivative of $n(\mathbf{r},t)\mathbf{v}(\mathbf{r},t)$:
\begin{flalign}
\frac{\partial[n(\mathbf{r},t)\mathbf{v}(\mathbf{r},t)]}{\partial t}    =\mathbf{v}(\mathbf{r},t)\frac{\partial n(\mathbf{r},t)}{\partial t}+n(\mathbf{r},t)\frac{\partial\mathbf{v}(\mathbf{r},t)}{\partial t}\nonumber \\
  =-\mathbf{v}(\mathbf{r},t)\bm{\nabla}\cdot[n(\mathbf{r},t)\mathbf{v}(\mathbf{r},t)]+n(\mathbf{r},t)\frac{\partial\mathbf{v}(\mathbf{r},t)}{\partial t}\nonumber \\
 =-\bm{\nabla}\cdot[n(\mathbf{r},t)\mathbf{v}(\mathbf{r},t)\mathbf{v}(\mathbf{r},t)]+n(\mathbf{r},t)[\mathbf{v}(\mathbf{r},t)\cdot\bm{\nabla}]\mathbf{v}(\mathbf{r},t) \nonumber\\+n(\mathbf{r},t)\frac{\partial\mathbf{v}(\mathbf{r},t)}{\partial t}.
\end{flalign}
Replacing the last term in Eq.~(\ref{eq:Euler_Wigner}) we find 
\begin{eqnarray}
n(\mathbf{r},t)\frac{\partial\mathbf{v}(\mathbf{r},t)}{\partial t}+n(\mathbf{r},t)[\mathbf{v}(\mathbf{r},t)\cdot\bm{\nabla}]\mathbf{v}(\mathbf{r},t)+\nonumber\\ +\bm{\nabla}\cdot\left[\bm{\Pi}-n(\mathbf{r},t)\mathbf{v}(\mathbf{r},t)\mathbf{v}(\mathbf{r},t)\right]={\cal \bm{P}_{W}}.
\end{eqnarray}
that results in the Euler equation.



%

\end{document}